\documentclass[prb,twocolumn,longbibliography,amsmath,floatfix,superscriptaddress,amssymb,nobibnotes]{revtex4-2}
\usepackage{tikz}
\usepackage{makeidx,amssymb,amsmath,amsthm,graphicx}
\usepackage[toc,page]{appendix}
\usepackage{dcolumn}
\usepackage{bm}
\usepackage{color}
\usepackage{placeins}
\usepackage{epstopdf}
\usepackage{multirow}
\usepackage{rcs}
\usepackage[normalem]{ulem}
\usepackage{chngcntr}
\usepackage[thinspace,thinqspace]{SIunits}
\usepackage{natbib}
\usepackage{hyperref}
\usepackage[hyphenbreaks]{breakurl}
\begin{document}
\title{Chemical pressure due to impurities in trigonal compounds Eu$T_2Pn_2$ ($T=$ Cd, Zn; $Pn=$ P, As, Sb)}
\author{Kristin~Kliemt}
\email[]{kliemt@physik.uni-frankfurt.de}
\affiliation{Physikalisches Institut, Goethe-Universit\"at Frankfurt/M, 60438 Frankfurt/M, Germany}
\date{\today}
%
\begin{abstract}
This work provides a review of crystal growth, crystal structure, compositional details, magnetism, thermodynamic, and transport behavior in the family of the trigonal intermetallic systems Eu$T_2Pn_2$ ($T=$ Cd, Zn; $Pn=$ P, As, Sb; space group $P\overline{3}m1$, No.164). The physical properties observed in these materials, and how these change depending on the growth conditions are discussed.  
In particular, the case of EuCd$_2$As$_2$ is considered where data from many sources are available. The possible small contamination of the material during crystal growth experiments is hard to verify as it is often below the detection limit of the standard characterization techniques. It turns out that samples from different sources exhibit variations in the lattice parameters exceeding the experimental errors.
The review of these parameters reveals that they are very similar for antiferromagnetic samples grown from Sn flux in Al$_2$O$_3$ crucibles, while there is a wider spread for samples grown from salt flux grown in SiO$_2$ ampules, which are mostly ferromagnetic. The influence of the different experimental setups with regard to possible impurities in the samples is discussed.

\end{abstract}
%
\maketitle
\def\neel{{N\'eel} }
\def\FA{F_{\rm an}}
\def\CA{C_{\rm an}}
\def\MS{M_{\rm sat}}
\def\EDF{E_{\rm df}  }
\def\text#1{{\rm #1}}
\def\i{\item}
\def\[{\begin{eqnarray*}}
\def\]{\end{eqnarray*}}
\def\bv{\begin{verbatim}}
\def\ev{\end{verbatim}}
\def\ganz{Z}
\def\3{\ss}
\def\reel{{\cal}R}
\def\platz{\;\;\;\;}
\def\beginvector{\left(\begin{array}{c}   }
\def\endvector{\end{array}\right)}
\def\fff{\frac{3}{k_B} F }
\def\vec#1{ {\rm \bf #1  } }
\def\KBMUEF{\frac{3k_B}{\mu_{\rm eff}^2 }}
%

\section{Introduction}

Solid state physics and materials science have always developed materials with potential for new technological applications like such showing extraordinary thermoelectric \cite{Zhu2015, Zeier2016} or transport properties being Weyl semimetals \cite{Wan2011, Xu2015, Armitage2018} or compounds exhibiting a colossal magnetoresistance (CMR) \cite{Ramirez1997}.  
Weyl semimetals, characterized by linear dispersion relations around certain points in the momentum space (Weyl points) where the conduction and valence bands touch, are a fascinating class of materials that have unique electronic properties due to their topological nature \cite{Armitage2018}.
These materials show a variety of interesting properties such as high carrier mobility, unusual magnetism, and robust surface states which make them promising candidates used in future technology like high-speed and low-power electronics, quantum computing or spintronics \cite{Sun2017, Venkateswara2019}. 
CMR materials where the electrical resistivity can be suppressed in field by several orders of magnitude \cite{Roeder1996, Chan1998, Lin2016, Rosa2020} are considered relevant to magnetic memory and sensing technologies \cite{HaghiriGosnet2003, Schneider2007}. 

\begin{figure*}
    \centering
    \includegraphics[width=0.85\linewidth]{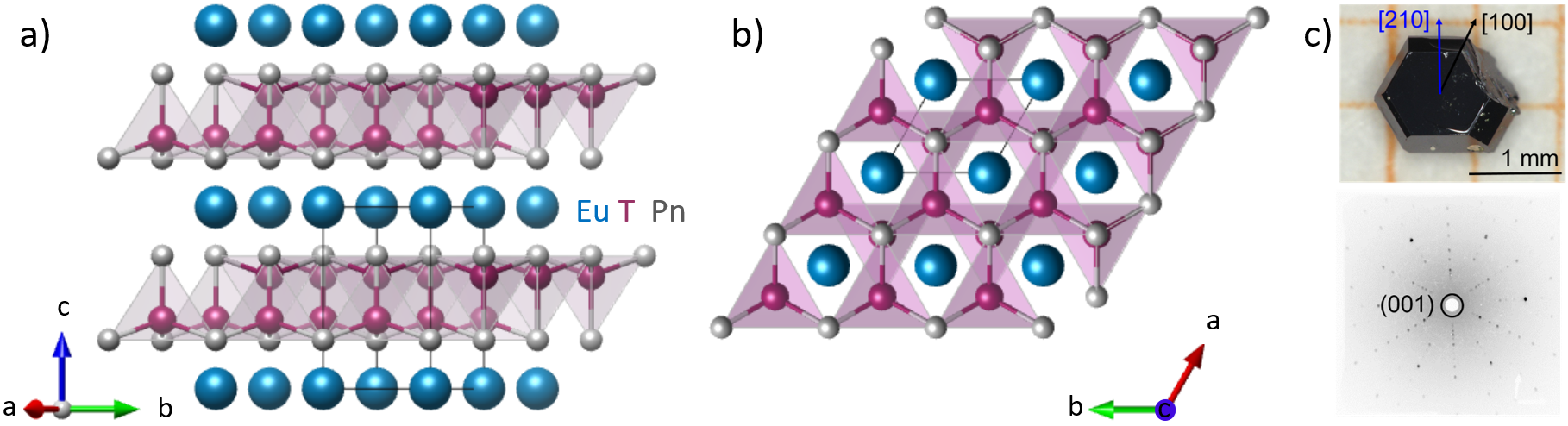}
    \caption{Eu$T_2Pn_2$ a) and b) trigonal CaAl$_2$Si$_2$-type crystal structure with the space group $P\overline{3}m1$ (No. 164). Solid gray lines show the unit cell. The picture was produced by a three-dimensional visualization system for crystallographic studies "VESTA"  \cite{Momma2011}.  c) Picture of a EuZn$_2$P$_2$ single crystal and Laue pattern (taken from \cite{Krebber2023}).}
    \label{fig:1}
\end{figure*}

The CaAl$_2$Si$_2$-structure type considered here, comprises $>400$ ternary compounds that are currently listed in the Inorganic Crystal Structure Database (ICSD) making this structure type  almost as versatile as the related ThCr$_2$Si$_2$-structure ($\approx 700$ entries) \cite{Shatruk2019}. While in the past, mostly Zintl phases with the CaAl$_2$Si$_2$ structure were known as promising thermoelectric materials \cite{Zheng2020, Peng2018, Shuai2017}, materials of this structure type again came into the focus of research with a proposal of CaMn$_2$Sb$_2$ being in proximity to a mean-field critical point known for the classical Heisenberg model on the honeycomb lattice \cite{Mazin2013, McNally2015}, the discovery of EuCd$_2$As$_2$ as a Weyl semimetal \cite{Ma2019} or EuCd$_2$P$_2$ as a CMR material \cite{Wang2021}. 
Especially the Eu-based compounds are of interest 
as these often exhibit strong exchange interaction between the large spin of the localized Eu$^{2+}$ ions and the charge carriers as well as small stray fields in case they order antiferromagnetically. 
In the recent past, the Eu$T_2Pn_2$ family with $T=$ Cd, Zn; and $Pn=$ P, As, Sb, crystallizing in the CaAl$_2$Si$_2$-type structure, Fig.~\ref{fig:1}, has attracted strong interest in the community as it comprises 
Weyl semimetals \cite{Ma2019,Luo2023}, CMR materials \cite{Wang2021, Du2022, Krebber2023} as well as  thermoelectrics \cite{Zhang2008, Zhang2010, Zhang2010a}.
It is known that the growth of Eu compounds poses special challenges due to the high tendency of Eu to oxidize and evaporate. Also, the samples properties might vary due to small shifts in the crystal`s stoichiometry \cite{Kliemt2022a}.
Therefore to identify a materials intrinsic properties, the quest for single crystals of high purity is immense.
In the semiconducting Eu$T_2Pn_2$ materials, small amounts of impurities can change the transport properties drastically e.g. induce metallicity \cite{Wang2021, Usachov2024, Zhang2023} and there is evidence that also the magnetic ground states can be influenced by small dopant, impurity or vacancy levels \cite{Jo2020} - the samples order either antiferromagnetic (AFM) or ferromagnetic (FM) depending on the growth conditions. 
Often, the origin of the sample dependencies is widely unclear since the differences in the samples like Eu deficiency or impurities are often close to or below the detection limit of the characterization techniques.
This review reveals that there are slight differences in the lattice parameters of samples from different sources which was not considered systematically so far.
This work provides an overview about published data of crystal growth conditions, structural analysis and physical properties of Eu$T_2$Pn$_2$ compounds ($T=$ Cd, Zn; $Pn=$ P, As, Sb). The samples that resulted from the different crystal growth procedures are compared with respect to their lattice parameters and their physical properties and possible connections are discussed.

\section{Crystal growth of Eu$T_2Pn_2$}
Trigonal Eu$T_2$Pn$_2$ compounds, space group $P\overline{3}m1$ (No.164), can be prepared in poly crystalline form by heating a stoichiometric ratio of the elements and a subsequent annealing step \cite{Kluefers1980, Artmann1996}. For the single crystal growth, in most cases a flux growth method using Sn, a NaCl/KCl 1:1 mixture or a self-flux (e.g. Sb or Zn-Sb) can be applied. Tabs.~\ref{Gitterkonstanten} and \ref{Gitterkonstanten2} provide an overview of the flux and crucible materials that were used, as well as the growth parameters and the results of the structural analysis by powder x-ray diffractometry (PXRD) or single crystal x-ray diffraction (SC XRD).
Some results of the chemical analysis are summarized in Tab.~\ref{EDX}. It is important to note that due to the large experimental error of the energy dispersive x-ray spectroscopy (EDX) analysis, it is not possible to determined a Eu deficiency by this method reliably. Instead, a single crystal x-ray analysis can be used to determine possible vacancies. 
Since Eu has a strong tendency to oxidize all handling and preparation steps need to be done in protective atmosphere in a glove box. During the growth, any contact with oxygen has to be avoided. Furthermore, the high vapor pressure of Eu at elevated temperature needs to be considered when designing crucible setups. 
Successful growth of single crystals was reported via external flux (for instance Sn in \cite{Singh2023,Berry2022,Wang2021,Wang2022}, NaCl-KCl  in \cite{Schellenberg2011,Jo2020,Chen2024_Cd,Chen2024_Zn}) or self-flux (Sb \cite{Singh2024}, Zn-Sb \cite{Weber2005, Weber2006}) in box furnaces or Bridgman setups. The removal of the flux was done via centrifugation or washing in water in case of salt flux after the crystal growth. 
In the crystal structure, Fig.~\ref{fig:1}a, Eu layers (blue) are separated by $T_2Pn_2$ blocks (red and gray). 
Fig.~\ref{fig:1}b depicts the view along $c$ onto the trigonal $a-b$ plane.
The Laue analysis of Sn-flux grown single crystals for EuZn$_2$P$_2$ \cite{Krebber2023}, and EuCd$_2$P$_2$ \cite{Usachov2024} confirms the orientation of the single crystals according to Fig.~\ref{fig:1}c.

\begin{table*}[bpth]
\begin{center}
\begin{tabular}{|c|cccccccccc|}
\hline\hline
Comp. & Flux/&molar ratio&Crucible&$T_{\rm max}/t_{\rm hold}$ &cool rate   &	$\quad a \quad$       &   $\quad c \quad$	&$c/a$   	 &  $\quad V \quad$& Ref.\\
         & method  &Eu:$T$:$Pn$:flux     &       &[$^{\circ}$C]$/$[h] &K/h           &[\AA]&[\AA]&&[\AA$^3$]&\\
\hline
\hline	
 &poly&1:2:2:0&Al$_2$O$_3$ &1000/24 & -&4.087(1) & 7.010(1)   &1.715 &101.411 &	\cite{Kluefers1980} \\
 &Sn	 &1:2:2:45  & Al$_2$O$_3$& 1150/24&5 & 4.08497(18)$^{\ddag}$	   & 7.0019(4)$^{\ddag}$ & 1.7141$^{\ddag}$&101.187(11)$^{\ddag}$ & \cite{Berry2022} \\
EuZn$_2$P$_2$& Sn	&1:2:2:20     & Al$_2$O$_3$   &1100/24& 2&4.0871&   7.0066      &1.7143&101.362(2)  & \cite{Krebber2023}\\
& Sn	& 1:2:2:45    & SiO$_2$   &1100/24&2&4.087&7.010 & 1.7152 &101.411(3) & \cite{Singh2023}\\
& Sn	& 1:1:1:8    & Al$_2$O$_3$&1100/24&3/5 &4.08685(2) &7.00784(5)& 1.7147 & 101.366(2)     &\cite{Chen2024_Zn}\\
& Sn	& \cite{Bukowski2022}    &Al$_2$O$_3$	 &1050/20&2-3 &4.0866(1) &7.0066(3)&1.7145  &101.339     &\cite{Rybicki2024}\\   
&&    & && &4.0779(1)$^{\diamond}$ &6.9834(3)$^{\diamond}$&1.7125$^{\diamond}$  & 100.561$^{\diamond}$    &\cite{Rybicki2024}\\  
\hline
EuZn$_2$P$_2$ & salt &  1:2:2$^{\dag\dag}$  &SiO$_2$ & 847/50 &1.5& 4.0765(2)$^{\dagger}$   & 6.9936(4)$^{\dagger}$  &1.7156$^{\dagger}$     & 100.648(11)$^{\dagger}$  & \cite{Chen2024_Zn}\\
\hline
\hline
 &poly&1:2:2:0&Al$_2$O$_3$ &1000/24 & -&4.211(1) & 7.181(1)   &1.705 &110.28 &	\cite{Kluefers1980}  \\
 &  poly  &1:2:2:0& Al$_2$O$_3$   &-	&-&  	-&- &-&-& \cite{Goryunov2012} \\
 &Sn &1:1:1:8&Al$_2$O$_3$  &1100/24&3/5  &4.21118(3) &	7.18114(6)    &	1.705 	& 110.2888(24)& \cite{Wang2022}\\
EuZn$_2$As$_2$ &Sn&1:2:2:20 &Al$_2$O$_3$	  &1000/10&3  &4.2093(1) &	7.175(3)   &	1.7046 	&110.09(6)&\cite{Blawat2022}\\
 &Sn&1:2:2:20 &Al$_2$O$_3$	  &1050/20&2-3&4.2119(6) &	7.1812(3)   &	1.7050	& 110.331(3)&\cite{Bukowski2022}\\
&Sn &\cite{Wang2022, Blawat2022,Bukowski2022}&Al$_2$O$_3$  &\cite{Wang2022, Blawat2022,Bukowski2022}&\cite{Wang2022, Blawat2022,Bukowski2022}& -&	-   &	-  	& -&\cite{Luo2023}\\
 &Sn &\cite{Wang2022, Blawat2022,Bukowski2022}&Al$_2$O$_3$ &\cite{Wang2022, Blawat2022,Bukowski2022}&\cite{Wang2022, Blawat2022,Bukowski2022}  &- &	 -  &	-  	& -&\cite{Yi2023}\\
\hline
 EuZn$_2$As$_2$  &salt&1:2:2$^{\dag\dag}$  &SiO$_2$ & 847/50 &1.5&-&-&-&-&\cite{Chen2024_Zn} \\
\hline
\hline
 &poly&1:2:2:0&Al$_2$O$_3$ &1000/24 & -&4.489(1) & 7.609(1)   &1.695 &132.7(7) &	\cite{Kluefers1980}  \\
 &Brid.&-&Al$_2$O$_3$&1150/48 & -&4.489&7.609 & 1.695   &132.78  &	\cite{Weber2005}  \\
 &Brid.&1:2:2:0&Al$_2$O$_3$&1150/10 & 10&4.489&7.609 & 1.695   &132.78  &	\cite{Weber2006}\\
EuZn$_2$Sb$_2$ &poly&1:2:2:0&C&900/336 & 33&4.4932(7)&7.6170(10)& 1.695   & 133.171 &	\cite{Zhang2008} \\ 
&  poly  &1:2:2:0 & C   &800/120&400&	4.4905(1)&7.6121(3)& 1.695 	& 132.93(1)&\cite{Zhang2010} \\
 &poly&1:2:2:0&Ta &1407/0.17 & -&4.4938(7) & 7.615(1)   &1.694(6) & 133.2   &	\cite{Schellenberg2010} \\
 &ZnSb&1:5:5:0&Al$_2$O$_3$&1000/12 &2 &4.4852(4)$^{\textasteriskcentered}$ &7.593(1)$^{\textasteriskcentered}$    & 1.6929$^{\textasteriskcentered}$&132.28(3)$^{\textasteriskcentered}$    &	\cite{May2012}  \\
 &Sb&1:2:45:0&Al$_2$O$_3$ &1000/20 &2 &- & -   & -& -   &	\cite{Singh2024} \\
\hline\hline
\end{tabular}
\end{center}
\caption{Room temperature lattice parameters $a$, $c$ ($^{\ddag}$at 213~K;  $^{\textasteriskcentered}$at 173~K;  $^{\dagger}$at 150~K; $^{\diamond}$at 15~K. Standard deviations in brackets.), $c/a$ and the volume $V$ of the unit cell of trigonal EuZn$_2Pn_2$ compounds, space group $P\overline{3}m1$ (No.164). Single crystals were grown from NaCl/KCl (salt, $^{\dag\dag}$weight ratio elements:salt = 1:4) or by Bridgman method (Brid.) and poly crystalline material (poly) was prepared without the usage of a flux.} \label{Gitterkonstanten}
\end{table*}

\begin{table*}[bpth]
\begin{center}
\begin{tabular}{|c|cccccccccc|}
\hline\hline
Comp. & Flux/&molar ratio&Crucible&$T_{\rm max}/t_{\rm hold}$ &cool rate   &	$\quad a \quad$       &   $\quad c \quad$	&$c/a$   	 &  $\quad V \quad$& Ref.\\
         & method  &Eu:$T$:$Pn$:flux     &       &[$^{\circ}$C]$/$[h] &K/h           &[\AA]&[\AA]&&[\AA$^3$]&\\
\hline
\hline	
  &poly   & 1:2:2:0&Al$_2$O$_3$& 850/25 & 	-   &	4.325(1)& 7.179(1) 	& 1.660&116.30&\cite{Artmann1996}\\
   &poly   &1($>3$N):2:2:0&Ta&822/100&quen.&4.3253(3)  &7.1833(4) 	   &	1.6608&116.4 &\cite{Schellenberg2011}\\
  &Sn   &1:2:2:20&Al$_2$O$_3$&950/36  & 	3   &	4.3248(2)& 7.1771(7) 	& 1.6595&116.26&\cite{Wang2021}\\
  & Sn & 1:2:2:20 &Al$_2$O$_3$&950/36  & 	3	  &	\cite{Wang2021}& \cite{Wang2021} 	&\cite{Wang2021}&\cite{Wang2021}&\cite{Sunko2023}\\
EuCd$_2$P$_2$  &Sn	& 1:2:2:20  &Al$_2$O$_3$&850/24  & 2	&4.324(1)   &	7.179(1)& 1.660(3)	 &116.25(2)& \cite{Usachov2024}\\
AFM  &Sn& 1:2:2:20  &C&850/24  & 2	&4.3178(5)&  7.1666(5)     &1.6598&115.711& \cite{Usachov2024}\\
  &Sn	   &1:1:1:20 \cite{Wang2021}&Al$_2$O$_3$&950/36  & 	3   &	-&  -	& -&-& \cite{Zhang2023}\\
   & Sn   &&Al$_2$O$_3$&950/36  & 	3 &4.3248(2)   &	7.1771(7)  	&1.6595&116.2451(5)&\cite{Chen2024_Cd}\\
\hline
EuCd$_2$P$_2$ FM &salt&  1:2:2$^{\dag\dag}$  &SiO$_2$&847/80  &1.5 	& 4.3168(2)$^{\dagger}$  &	7.1616(5)$^{\dagger}$  	& 1.6590$^{\dagger}$&115.575(13)$^{\dagger}$&\cite{Chen2024_Cd}\\
\hline\hline
  &poly   &1:2:2:0 &Al$_2$O$_3$& 850/25 & 	-   &	4.439(1)& 7.328(1)	& 1.651&125.04 &\cite{Artmann1996}\\
& poly &1($>3$N):2:2 &Ta& 822/100   &	quench.&4.4499(9)&7.350(1)  	& 1.651689&126.0 &\cite{Schellenberg2011} \\  
 &	Sn & 1:2:2:10&Al$_2$O$_3$&900/20   & 2 	&-   &	-  	& -&-& \cite{Ma2019} \\
 &	Sn & 1:2:2:10&Al$_2$O$_3$&900/20   & 2& -  &	-  	& -&-&\cite{Ma2020} \\
 &	Sn & 1:2:2:10 &Al$_2$O$_3$&900/20   & 2.5& 4.4412  &7.3348 	&1.6515 &125.291&\cite{Jo2020}\cite{Gati2021} \\
EuCd$_2$As$_2$  &	Sn & 1:2:2:10 &Al$_2$O$_3$&900/20   & 2& -  &	7.33 	&- &125.05&\cite{Sun2022}\cite{Du2022} \\
AFM&	Sn &1:2:2:10&Al$_2$O$_3$  &900/20&2  &4.43645   &	7.32004  	& 1.64998& 124.782&\cite{Cao2022} \\
&	Sn &1(2N):2:2:10&Al$_2$O$_3$  &900/20&\cite{Ma2019} 2  &4.43995(6)$^{*}$   &	7.3274(2)$^{*}$  	& 1.65034$^{*}$& 125.103$^{*}$&\cite{Santos2023}\\
&	Sn &1(4N):2:2:10&Al$_2$O$_3$  &900/20&\cite{Ma2019} 2  &4.43956(6)$^{**}$   &	7.3266(1)$^{**}$  	& 1.65030$^{**}$& 125.065$^{**}$&\cite{Santos2023}\\
&	Bi &-&- &-&-  &4.43401$^{\diamond}$   &	7.31337$^{\diamond}$  	& 1.6494$^{\diamond}$& 124.522$^{\diamond}$&\cite{Chen2023_CdAs}\\
& Sn&1:2:2:10 &Al$_2$O$_3$&900/20&2&4.4482(3)&7.3413(8)&1.65040&125.81&\cite{Roychowdhury2023}\\
&	Sn &1:2:2:10\cite{Jo2020}&Al$_2$O$_3$ &900/24&  1.75&4.4386$^{**}$   &	7.3239$^{**}$  	&1.65005$^{**}$& 124.970$^{**}$&\cite{Shi2024}\\
&	Sn &1(4N):2:2:10+La&Al$_2$O$_3$ &900/20&  2&4.44012(9)   &	7.32664(10)  	&1.6501& 125.0907&\cite{Nelson2024}\\
&	Sn &1(4N):2:2:10&Al$_2$O$_3$ &900/20&  2&4.43979(5)   &	7.32690(6)  	&1.65028& 125.0765&\cite{Nelson2024}\\
\hline
& salt &1($>3$N):2:2 &SiO$_2$& 847/100   &	5&-&-  	& -&-&\cite{Schellenberg2011}\\ 
& salt &\cite{Schellenberg2011} &SiO$_2$& 847/100   &	5&-&7.34  	& -&- &\cite{Wang2016} \\
EuCd$_2$As$_2$& salt &\cite{Schellenberg2011} &SiO$_2$&  847/100   &	5&$\approx 4.44$& $\approx 7.33$ 	& $\approx 1.65$&$\approx 125.15$ &\cite{Rahn2018} \\
AFM &	salt & 1.75:2:2 &SiO$_2$&847   & 5	& 4.4398(2)  	& 7.3277(4)& 1.6505&125.084 &\cite{Jo2020}\\
 &	 &&&   &	& 4.43219(1)$^{\$}$  	& 7.30436(1)$^{\$}$&1.64803$^{\$}$ &124.2715$^{\$}$ &\cite{Jo2020}  \\
\hline
 &	salt & 1:2:2 &SiO$_2$&847   & 5	& 4.4365(2)  &	7.3247(4)	&1.6510&124.862&\cite{Jo2020}\\
EuCd$_2$As$_2$ &	 &&&   &	& 4.43021(1)$^{\$}$  	& 7.30240(2)$^{\$}$&1.64832$^{\$}$ &124.117$^{\$}$ &\cite{Jo2020}  \\
FM&	salt &1:2:2 &SiO$_2$&847/100   &	1&   4.4412(2)& 7.3255(8)&1.64944 & 125.13(2)&\cite{Sanjeewa2020}\\
& salt&1:2:2:4:4 &Al$_2$O$_3$&847/100&1&4.4437(4)&7.3313(11)&1.64981&125.36523&\cite{Roychowdhury2023}\\
\hline
\hline
	 &  poly &1:2:2:0&Al$_2$O$_3$&850/10  &  - &	4.698(1)&7.723(1) 	&1.644& 147.612  &\cite{Artmann1996}\\
 &  poly  &1:2:2:0 & C   &800/120&400&	4.6991(1)&7.7256(2)&1.644  	&147.74(1) &\cite{Zhang2010} \\
 &poly&1:2:2:0 &C-SiO$_2$       &850/120&-&	-&-&-  	&-&\cite{Zhang2010a} \\
&  poly  &1:2:2:0& C   &1200/24	&ann.&4.6991(1)  	&7.7256(2) &1.644&147.74(1) &\cite{Zhang2010b} \\
EuCd$_2$Sb$_2$  &poly & 1:2:2:0 &Ta&1407/0.17&ann.&4.699(2)  &7.725(2)	   &	1.64396&147.7 &\cite{Schellenberg2011}\\
&  poly  &1:2:2:0 &Al$_2$O$_3$   &-	&-&-  	&- &-&- &\cite{Goryunov2012} \\
 & VT/I   &1:2:2:x& Al$_2$O$_3$   &	1052/168&slow&4.7030(9) 	& 7.7267(18)&1.64293&147.99206 &\cite{Soh2018} \\
 &  \cite{Soh2018}  & \cite{Soh2018}&Al$_2$O$_3$   &	1052/168&slow&4.6926(6)&7.7072(8)  	 &1.642415&146.9763&\cite{Su2020} \\
\hline
EuCd$_2$Sb$_2$ &  poly & 1:2.2:0 & Al$_2$O$_3$   &	1000/24&&4.4639(3)  	&5.6511(5) &1.265955&112.61(3)&\cite{Jimenez2023} \\
$P4/mmm$       &  600$^{\circ}$C&6GPa  &     &	    	& &             &  &&& \\
\hline\hline
\end{tabular}
\end{center}
\caption{Room temperature lattice parameters $a$, $c$ ($^{\dagger}$at 150~K; $^{\$}$at 100~K; $^{\diamond}$under 0.5\,GPa. Standard deviations in brackets.), $c/a$ and the volume $V$ of the unit cell of trigonal EuCd$_2Pn_2$ compounds, space group $P\overline{3}m1$ (No.164). Single crystals were grown in NaCl/KCl (salt, $^{\dag\dag}$weight ratio elements:salt = 1:4.), by Bridgman method (Brid.) or vapour transport (VT), poly crystalline material (poly) was prepared without the usage of a flux. Resulting samples were $^{*}$metallic or $^{**}$insulating. }
\label{Gitterkonstanten2}
\end{table*}

\begin{table}[bpth]
\begin{center}
\begin{tabular}{|c|cccc|ccccccc|}
\hline\hline
Comp.        & Eu         &$T$      &$Pn$        &Ref. \\
             & [at.$\%$]  &[at$\%$] & [at$\%$]&           \\
\hline
\hline	
EuZn$_2$P$_2$& $21\pm2$   &$38\pm2$    & $40\pm2$   &      \cite{Krebber2023}           \\
AFM, Sn & 21.11 & 39.03   & 39.87  &  \cite{Singh2023} \\
             & $19.8\pm0.4$& 40    & $40\pm3 $& \cite{Chen2024_Zn} \\
             &19.4&39.6    &41.0 &\cite{Rybicki2024}\\     
\hline
EuZn$_2$P$_2$  &$19.2\pm 0.6$$^{\dag\dag}$ & 40&$40.8\pm 1.2$ &\cite{Chen2024_Zn}  \\
salt, FM&&&&\\
\hline
\hline
EuZn$_2$As$_2$ &$22.5\pm0.2$&$39.8\pm0.4$ &$37.2\pm0.2$  &\cite{Blawat2022}\\
AFM, Sn&$20\pm1$&$40\pm2$ &$40\pm2$	  &\cite{Bukowski2022}\\
&19.65 &40.04&40.31  &\cite{Luo2023}\\
 &- &-&- &\cite{Yi2023} \\
 \hline
 EuZn$_2$As$_2$  &$18.7\pm0.5$ & $41.6\pm1.2$ &$39.7\pm0.7$ &\cite{Chen2024_Zn} \\
 salt, FM&&&&\\
\hline
\hline
EuZn$_2$Sb$_2$  &$19.6\pm0.2$&$39.6\pm0.2 $&$40.2\pm0.2$ &\cite{Zhang2010}\\
 &21.92&38.94&39.13 &\cite{Singh2024}\\
\hline
\hline
EuCd$_2$P$_2$   &20   &40.04&40.06&\cite{Sunko2023}  \\
AFM, Sn &$17\pm3$	& $40\pm2$  &$43\pm3$&\cite{Usachov2024}  \\
\hline
EuCd$_2$P$_2$&$19.90\pm0.06$$^{\dag}$&  40$^{\dag}$  &40$^{\dag}$&\cite{Chen2024_Cd} \\
salt&    &&&  \\
\hline\hline
EuCd$_2$As$_2$  &	20$^{\dag}$ & - &-&\cite{Jo2020}   \\
 salt, AFM&&&&\\
EuCd$_2$As$_2$ &	$19.2\pm0.2$$^{\dag}$$^{\textasteriskcentered}$ & - &-&\cite{Jo2020}   \\
 salt, FM&&&&\\
EuCd$_2$As$_2$&	$19.9\pm0.2$$^{\textasteriskcentered\textasteriskcentered}$ & $39.6\pm0.2$ &$40.4\pm0.2$&\cite{Sanjeewa2020}   \\
 salt, FM&&&&\\
\hline
\hline
EuCd$_2$Sb$_2$ &  $18.8\pm0.6$  &$41.4\pm0.8$ & $39.8\pm1.4$   &\cite{Zhang2010, Zhang2010b} \\
\hline
\hline\hline
\end{tabular}
\end{center}
\caption{Results of energy dispersive x-ray spectroscopy on trigonal Eu$T_2Pn_2$ compounds space group $P\overline{3}m1$ (No.164). $^{\dag}$site occupancy from single crystal analysis. $^{\dag\dag}$$5\%$ Eu vacancies from SC XRD, $^{\textasteriskcentered}$Eu deficit, $^{\textasteriskcentered\textasteriskcentered}$no Eu deficit. The compositional details for p- and n-type EuCd$_2$As$_2$ crystals were studied by time-of-flight secondary-ion-mass spectroscopy (ToF-SIMS) in \cite{Nelson2024}. \label{EDX}}
\end{table}

\section{Magnetic properties}
In the Eu$T_2Pn_2$ compounds, Eu is in a stable 2+ configuration and with $L=0$, the total angular momentum equals the large pure spin momentum $J=S=7/2$ and minimal crystalline electric field effects are expected at most. 
In $M(H)$ measurements, the saturation is reached at $M_{\rm sat}=g_JJ=7\mu_B$/Eu with $g_J$ being the Land\'{e} factor. The calculated value of the effective magnetic moment is $\mu_{\rm eff}^{\rm calc}=\sqrt{J(J+1)}\mu_B=7.94\,\mu_B$.

\begin{figure}[htbp]
    \centering
    \includegraphics[width=1.0\linewidth]{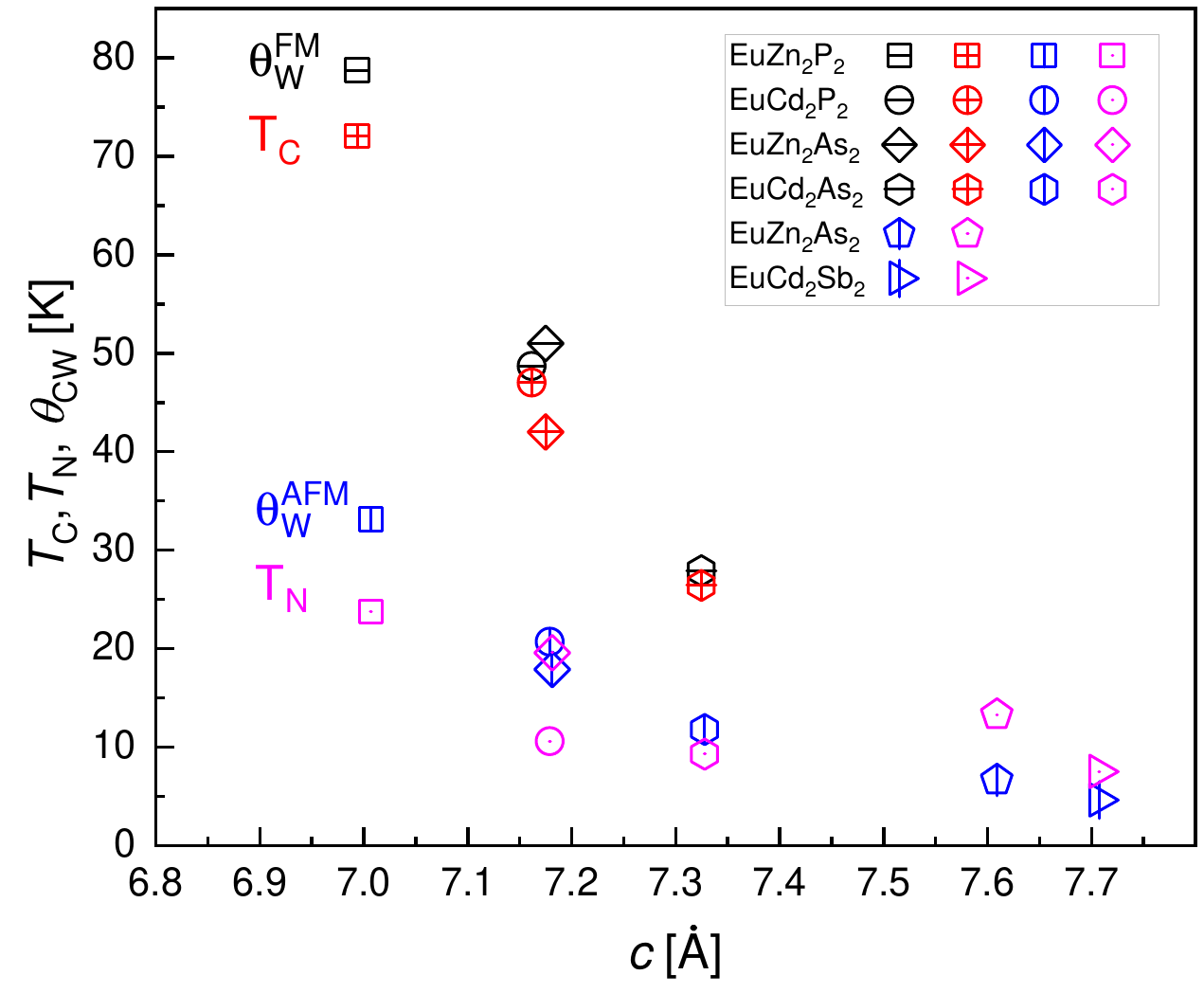}
    \caption{$T_C, T_N$ and $\Theta_W$ as a function of the $c$ lattice parameter for  EuZn$_2$P$_2$ (squares) \cite{Chen2024_Zn, Krebber2023}, EuCd$_2$P$_2$ (circles) \cite{Chen2024_Cd, Usachov2024}, EuZn$_2$As$_2$ (diamonds) \cite{Chen2024_Zn, Wang2022}, and EuCd$_2$As$_2$ (hexagons) \cite{Jo2020, Artmann1996}, EuZn$_2$Sb$_2$ (pentagons) \cite{Weber2006}, and EuCd$_2$Sb$_2$ (triangles) \cite{Su2020}. Uncertainties of the data are smaller that the sizes of the symbols. Figure after \cite{Chen2024_Zn}.}
    \label{fig:8}
\end{figure}

The FM-Eu$T_2Pn_2$, Fig.~\ref{fig:8}, show a linear relationship between the Weiss temperatures $\Theta_W$ (the Curie temperatures $T_C$) and the $c$ lattice parameter (distance of the Eu layers) which suggests that $T_C$ can be enhanced by increasing the Eu layer distance  \cite{Chen2024_Zn}. It was proposed that EuCd$_2$As$_2$ is a  candidate for a magnetic Weyl semimetal if it orders FM with the magnetic moments aligned parallel to the crystallographic $c$ axis \cite{Wang2019} meaning that the exact experimental verification of its ground state magnetic structure as well as those of other Eu$T_2Pn_2$ is of central importance.

The determination of the magnetic structure of Eu compounds via neutron scattering is often difficult due to the high absorption cross section of Eu. Furthermore, in the Eu$T_2Pn_2$ family, where physical properties are sensitive to small impurity levels even using isotope pure material to grow less absorbing crystals is problematic as its purity is limited as for instance to Eu (2N) in \cite{Taddei2022}.
Alternative ways to get information about the magnetic structure of rare earth compounds even with low magnetic anisotropy have been proven to be low field magnetization \cite{Kliemt2017, Pakhira2022, Kliemt2023, Krebber2023}, electron spin resonance \cite{Sichelschmidt2018}, resonant elastic x-ray scattering (REXS) \cite{Rahn2018}, or photoemission \cite{Usachov2022, Usachov2023} which might be applied to the Eu$T_2Pn_2$ materials in future studies. 

\section{Thermodynamic properties}

The magnetic contribution to the specific heat can be obtained by subtracting the phonon contribution and the measured specific heat data can be fitted using 
\begin{equation}
    C_p(T) = \gamma T + (1-m)C_D(T) + mC_E(T), \label{HCFIT}
\end{equation}
where $m$ is weight of the Einstein term and $\gamma$ is the Sommerfeld coefficient. Here, 
\begin{equation}
C_D(T)=9nR\left(\frac{T}{\Theta_D}\right)^3 \int_0^{\frac{\Theta_D}{T}} \frac{x^4{\rm e}^x}{{(\rm e}^{x}-1)^2} {\rm d}x,
\end{equation} 
and
\begin{equation}
C_E(T)=3nR\left(\frac{\Theta_E}{T}\right)^2  \frac{{\rm e}^{\Theta_E/T}}{({\rm e}^{\Theta_E/T}-1)^2} 
\end{equation} 
 where $n$ is the number of atoms per formula unit, $R$ is the universal gas constant, $\Theta_D$ and $\Theta_E$ are Debye
temperature and Einstein temperature, respectively (see e.g. \cite{Singh2024}).
Through integration of the heat capacity, the magnetic part of the entropy can be obtained. The expected value for Eu$^{2+}$ with J$=7/2$ is $S_{\rm mag}=\rm{R\,ln}(2J+1)=17.28\rm\, Jmol^{-1}K^{-1}$.

\section{Transport properties}
The Eu$T_2Pn_2$ compounds are bad conductors and special attention has to be paid to the preparation of the contacts for transport measurements as for instance without polishing the crystal surface and depositing a Cr/Au layer, non-linear U(I) curves might be obtained \cite{Krebber2023, Usachov2024}.
Throughout literature, two different definitions are used to quantify the magnetoresistance (MR)

\begin{equation}
\text{MR} = 100\% \times [\rho(H)-\rho(0)]/\rho(H) \label{MR1}
\end{equation} 
or
\begin{equation}
\text{MR} = 100\% \times [\rho(H)-\rho(0)]/\rho(0).\label{MR2}
\end{equation}
Note that using Eqn.~\ref{MR1}, values larger than $100\%$ can be obtained, while MR values calculated using Eqn.~\ref{MR2} do not exceed $100\%$.
The general expression for the total Hall resistivity is
\begin{equation}
    \rho_{xy}= R_0\mu_0H + R_{S}M + \rho^{NL}_{xy},
    \label{HallRes}
\end{equation}
where $R_0\mu_0H$ is linear in the field ($\mu_0H$) and represents the ordinary Hall effect (OHE),
while the second term $R_{S}M$ is linear in the magnetization ($M$) and represents the conventional anomalous Hall effect (AHE). 
Furthermore, the third term, $\rho^{NL}_{xy}$, is not linear in $\mu_0H$ or $M$ and represents the nonlinear AHE (NLAHE), which is characterized by  peaks in $\rho_{xy}(H)$ curves. 

\section{Materials}
\begin{figure*}[htbp]
    \centering
    \includegraphics[width=1.0\textwidth]{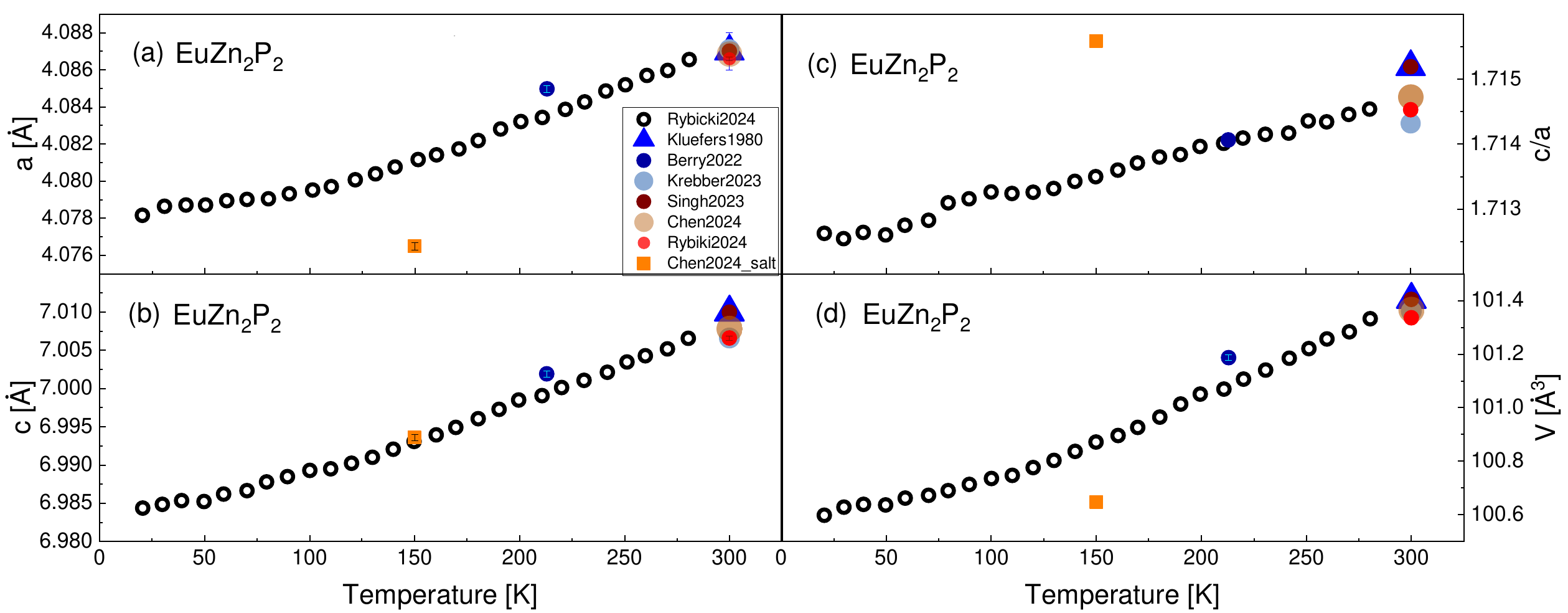}
    \caption{EuZn$_2$P$_2$: Temperature dependence of the lattice parameters \cite{Rybicki2024} (open symbols), of poly crystalline samples (triangles) \cite{Kluefers1980} as well as of single crystals grown from Sn flux \cite{Berry2022, Krebber2023, Singh2023, Chen2024_Zn, Rybicki2024} (closed circles). The orange squares indicate the values measured on samples grown from salt flux \cite{Chen2024_Zn}. (c) $c/a$ ratio and (d) the volume $V$ of the unit cell of the same samples. }
    \label{fig:2}
\end{figure*}

\begin{figure*}[htbp]
    \centering
    \includegraphics[width=1.0\textwidth]{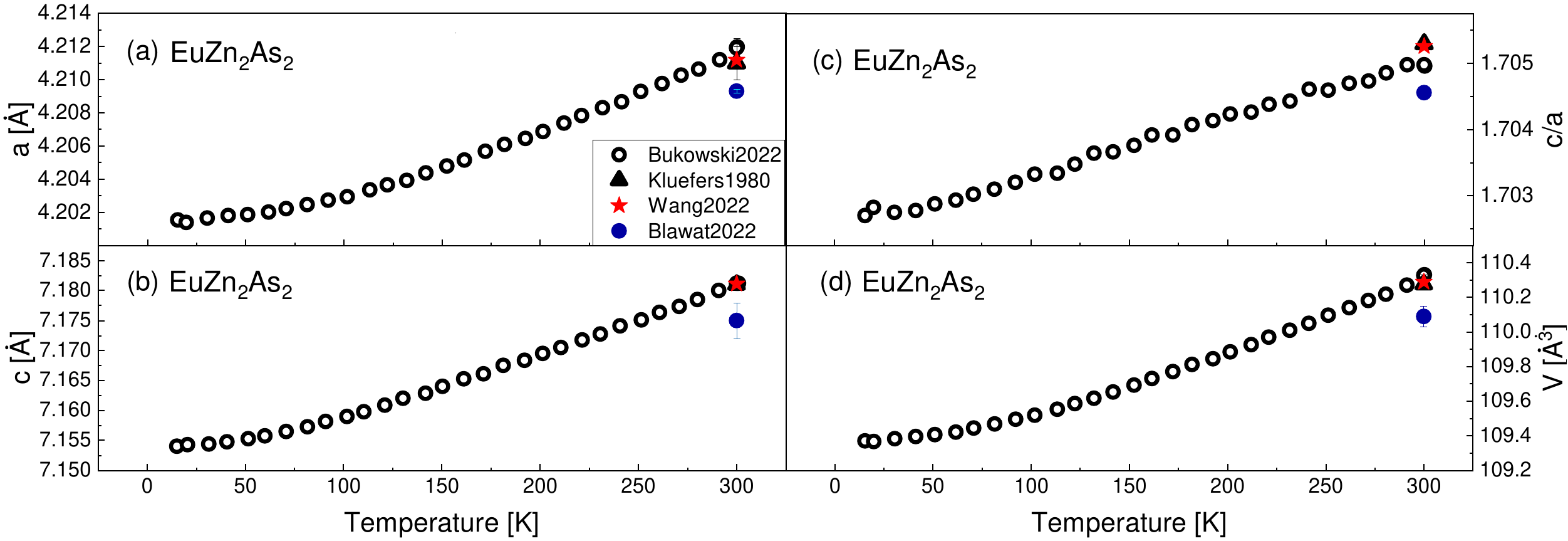}
    \caption{EuZn$_2$As$_2$: Temperature dependence of the lattice parameters \cite{Bukowski2022} (open circles), of poly crystalline samples (triangles) \cite{Kluefers1980} as well as of single crystals grown from Sn flux \cite{Blawat2022} (closed circles). The asterisk indicate the values measured on samples grown using an Eu excess in the Sn flux \cite{Wang2022}. (c) $c/a$ ratio and (d) the volume $V$ of the unit cell of the same samples. } 
    \label{fig:3}
\end{figure*}

\subsection{EuZn$_2$P$_2$ }
\subsubsection{Crystal growth, structural and chemical   characterization}
In the case of EuZn$_2$P$_2$, poly crystalline material was prepared in Al$_2$O$_3$ crucibles \cite{Kluefers1980} and single crystals were grown from tin flux using Al$_2$O$_3$, C or SiO$_2$ \cite{Berry2022, Krebber2023, Singh2023, Chen2024_Zn, Rybicki2024} crucibles or from salt flux in SiO$_2$ \cite{Chen2024_Zn}.
The temperature dependence of the lattice parameters of EuZn$_2$P$_2$ (open symbols, data taken from \cite{Rybicki2024}) is shown in Fig.~\ref{fig:2}. 
 The $a$ and $c$ parameters, as well as the  $c/a$ ratio and the volume of the unit cell from different sources are depicted for comparison.
While poly crystalline samples (triangles)  \cite{Kluefers1980} as well as single crystals grown from Sn flux (closed circles) \cite{Berry2022, Krebber2023, Singh2023, Chen2024_Zn, Rybicki2024}, show similar lattice parameters, the $a$ lattice parameter of single crystals grown from salt flux (closed squares) \cite{Chen2024_Zn} is $\approx 0.11\%$ smaller while the $c$ parameter matches that of a sample grown from Sn flux \cite{Rybicki2024}. Therefore, the volume of the unit cell in case of salt flux usage is 0.22\% smaller (at 150~K) than in case of Sn flux. An analysis by single crystal x-ray diffraction reveals that the samples exhibit 5\% Eu vacancies \cite{Chen2024_Zn}.  Energy dispersive x-ray spectroscopy, Tab.~\ref{EDX}, revealed an elemental ratio of Eu:Zn:P =$(19.2\pm 0.6) : 40 : (40.8\pm1.2)$ and supports the SC-XRD result that samples grown from salt flux (SiO$_2$ crucible) exhibit an Eu deficit, while the elemental ratio is close to the ideal composition Eu:Zn:P =$(19.8\pm 0.4) : 40 : (40\pm3)$ for Sn-flux-grown samples (Al$_2$O$_3$ crucible) \cite{Chen2024_Zn}.
Interestingly samples grown from Sn flux (SiO$_2$ \cite{Singh2023} or C \cite{Krebber2023} crucible) show a slight Eu excess in the EDX analysis. During our work \cite{Krebber2023} we could not find indications in the EDX spectra for Al from Al$_2$O$_3$ crucible  usage being incorporated in the crystals. Nevertheless, we observed oxidation of Eu when using Al$_2$O$_3$ crucibles and changed the crucible material to C. All samples used in the study \cite{Krebber2023} for the physical characterization were grown from C crucibles.
From the temperature dependence of the unit cell volume, Fig.~\ref{fig:2}d, the linear thermal expansion coefficients 
$\alpha_a=1.01\cdot 10^{-5}\,\rm K^{-1}$, $\alpha_c=1.48\cdot 10^{-5}\,\rm K^{-1}$ and the volume thermal expansion $\alpha_V=3.50\cdot 10^{-5}\,\rm K^{-1}$ \cite{Rybicki2024} of EuZn$_2$P$_2$ (Sn flux) were determined using the Debye formula \cite{Sayetat1998}
\begin{equation}
V= V_0+I_C\frac{T^4}{\Theta_D^3}  \int_0^{\frac{\Theta_D}{T}} \frac{x}{{\rm e}^{x}-1} dx,
\end{equation} 
where $V_0$ is the unit cell volume extrapolated to 0~K, $I_C$ is the 
slope of the linear part of $V(T)$ dependent on Gr\"uneisen and compressibility parameters, while $\Theta_D$ is the Debye temperature
(for details see \cite{Rybicki2024}).

\subsubsection{Magnetic and thermodynamik properties (poly, Sn flux)}

In EuZn$_2$P$_2$ grown from Sn flux \cite{Berry2022, Krebber2023, Singh2023,Chen2024_Zn,Rybicki2024} the Eu$^{2+}$ ions order AFM below $T_{\rm N}\approx 23.5\,\rm K$. A summary of some magnetic propeties can be found in Tab.~\ref{Tabellemagnet_Zn}. In Refs.~\cite{Berry2022,Singh2023} an A-type, AFM structure with magnetic moments aligned in the $a-a$ plane was proposed.
Soon after, resonant magnetic diffraction confirmed the simple "+ - + -" AFM stacking of FM ordered Eu planes (A-type AFM structure, $q_{AF}=(0\, 0\, 1/2)$) experimentally but revealed an alignment of the magnetic moments along the Eu-Eu direction in the $a-a$ plane and a canting by $40^{\circ}\pm 10^{\circ}$ away from the $a-a$ plane instead \cite{Krebber2023}. The latter was confirmed through $^{151}$Eu M\"ossbauer spectroscopy in Ref.~\cite{Rybicki2024}.
EuZn$_2$P$_2$ shows semiconducting behaviour and the highest $T_{\rm N}$ among EuZn$_2X_2$ compounds (where $X$=P, As, Sb, see \cite{Berry2022} and Fig.~\ref{fig:8}). 
It was argued that this high $T_{\rm N}$ is due to dipolar interactions upon absence of conduction electrons needed for the Ruderman–Kittel–Kasuya–Yosida (RKKY) exchange interaction \cite{Berry2022}.
In a more recent work, the magnetism was explained by weak antiferromagnetic interactions along the $c$ axis and a strong ferromagnetic Eu-Eu exchange coupling within the basal plane sheets, based on superexchange including phosphorus ions \cite{Singh2023}. 
Magnetic order at $T_{\rm N}\approx23.5\,\rm K$ was confirmed through heat capacity measurements \cite{Berry2022, Krebber2023, Singh2023, Rybicki2024} but the different studies use different models to determine the Debye temperature and the magnetic part of the entropy  $S_{\rm mag}$. Using a model, similar to Eqn.~\ref{HCFIT}, which comprises a Debye temperature $\Theta_D=(277\pm8)\,\rm K$ and two Einstein temperatures $\Theta_{E1}=(130\pm1)\,\rm K$ and $\Theta_{E2}=(406\pm2)\,\rm K$ describes the experimental data well and enabled the determination of $S_{\rm mag}=16.4\rm\, Jmol^{-1}K^{-1}$ \cite{Rybicki2024}.
This is quite close to the expected value for Eu$^{2+}$ with J$=7/2$, $S_{\rm mag}=\rm{R\,ln}(2J+1)=17.28\rm\, Jmol^{-1}K^{-1}$.

\subsubsection{Transport and electronic properties (poly, Sn flux)}
In first studies, EuZn$_2$P$_2$ was investigated down to $\approx 140\,\rm K$ and insulating behaviour was reported \cite{Berry2022, Singh2023}.
In a later work \cite{Krebber2023}, the material was studied down to low temperatures and a colossal magnetoresistance effect (CMR) above and below $T_{\rm N}$ was found which was confirmed later by \cite{Rybicki2024} through resistance measurements. 
It has to be noted, that the reported temperature dependence of resistivity and resistance is different in that sense, that in the resistivity \cite{Krebber2023} data a kink at $T_N$ appears, while the resistance \cite{Rybicki2024} shows a maximum above the \neel temperature which shifts up to higher T with increasing field or pressure. 
There, it was discussed that the CMR might be related to a reduction of the band-gap width and that it can be suppressed in a pressure of  $p\approx 17\,\rm GPa$. Furthermore, it was reported that hydrostatic pressure increases the \neel temperature up to $T^{10\,{\rm GPa}}_{\rm N}=45\,\rm K$ \cite{Rybicki2024}. 
The $T$ dependence of the resistance of EuZn$_2$P$_2$ reported in \cite{Rybicki2024} shows similarities to the transport behaviour of EuCd$_2$P$_2$ where a peak in the resistivity is observed above $T_N$ which is discussed to originate from the formation and percolation of magnetic polarons \cite{Wang2021}. 
In EuZn$_2$P$_2$, a considerable negative magnetoresistance (MR) of order a few \% is observed even for small fields and already at temperatures as high as $\approx 150\,\rm K$. The MR becomes stronger in higher magnetic fields when lowering the temperature down to $T_{\rm N}$. To calculate the MR, the definition according to Eqn.~(\ref{MR2}) was used.
The strong negative MR reaches saturation values of $-99.4$\,\% at $\mu_0H = 4$\,T and $-99.9$\,\% at $11$\,T \cite{Krebber2023}.
Several attempts were reported to determine the band gap of EuZn$_2$P$_2$ experimentally (e.g. 0.11\,eV \cite{Berry2022}, $\approx 0.35$\,eV ($\approx0.26$\,eV at $\mu_0H=14\,\rm T$)\cite{Singh2023}, 0.29\,eV \cite{Rybicki2024} from an Arrhenius plot of the electrical transport data) or theoretically using DFT+U \cite{Berry2022, Singh2023} but here the authors show that these values of the indirect and direct gaps obviously strongly depend on the treatment of the $4f$ electrons and the correlation parameter $U$ (see for instance Table S1 in the supplemental material of \cite{Singh2023}).
Recently, the indirect and direct gaps were determined in a combined study using ARPES, optical spectroscopy and DFT+U calculations. An indirect gap of ${E_i^\mathrm{DFT}=0.2\,\rm eV}$ and a direct band gap of ${E_d^\mathrm{DFT}=0.34\,\rm eV}$ were predicted theoretically which corresponds very well with the experimentally determined band gaps $E_i^\mathrm{opt}=0.09\,\rm eV$ and $E_d^\mathrm{opt}=0.33\,\rm eV$ \cite{Krebber2023}.

\subsubsection{Magnetic properties (salt flux)}

In EuZn$_2$P$_2$ grown from salt flux  ferromagnetic order is observed below $T_{\rm C}=72\,\rm K$ \cite{Chen2024_Zn}. For comparison with the AFM material, some magnetic properties are listed in Tab.~\ref{Tabellemagnet_Zn}. 
Single crystal diffraction revealed that ferromagnetic EuZn$_2$P$_2$ exhibits Eu vacancies ($\approx 5\%$) which lead to a heavy hole doping. It was proposed that these induced additional carriers are responsible for an interlayer FM coupling, which results in the transition from insulating AFM EuZn$_2$P$_2$ to metallic FM EuZn$_2$P$_2$ \cite{Chen2024_Zn}. 
According to the $M(H)$ curves, ferromagnetic EuZn$_2$P$_2$ exhibits a strong magnetocrystalline anisotropy and from the occurrence of a hysteresis in $M(H)$ for $H\perp001$ it is deduced that the magnetic moments are aligned in the $a-a$ plane \cite{Chen2024_Zn}.

\subsubsection{Transport and electronic properties (salt flux)}
In contrast to semiconducting AFM-EuZn$_2$P$_2$ \cite{Krebber2023}, FM-EuZn$_2$P$_2$ exhibits metallic behavior with residual resistivity of $\approx 0.2\,\rm m \Omega cm$ at $2\,\rm K$ \cite{Chen2024_Zn}.
Based on the single-band model, the hole concentration in FM-EuZn$_2$P$_2$ was estimated to be 10$^{21}$cm$^{-3}$ by considering $5\%$ Eu$^{2+}$ vacancies in the lattice \cite{Chen2024_Zn}.
At $T_C$, the maximum negative MR is achieved, which amounts to over $-50\%$ at 8 T  \cite{Chen2024_Zn}.

\subsection{EuZn$_2$As$_2$}
\subsubsection{Crystal growth, structural and chemical characterization}
In the case of EuZn$_2$As$_2$, poly crystalline material was prepared in Al$_2$O$_3$ crucibles \cite{Kluefers1980} and single crystals were grown from tin flux using Al$_2$O$_3$ crucibles  \cite{Wang2022, Bukowski2022, Blawat2022, Luo2023, Yi2023} or from salt flux in SiO$_2$ \cite{Chen2024_Zn}.
The structural parameters for the different kinds of preparation  \cite{Kluefers1980, Wang2022, Bukowski2022, Blawat2022} are compared in Fig.~\ref{fig:3}. No lattice parameters are given for the samples investigated in \cite{Goryunov2012, Luo2023, Yi2023} as well as for the salt-flux grown samples \cite{Chen2024_Zn}. Poly crystalline as well as samples grown from Sn flux have nearly identical lattice parameters.
The chemical analysis reveals that samples grown from Sn flux in Al$_2$O$_3$ crucibles show a stoichiometry close to the ideal 122 composition \cite{Blawat2022, Bukowski2022}. It can be speculated that an Eu excess in the sample \cite{Luo2023} was most likely achieved by using an Eu excess in the initial melt \cite{Wang2022}.

\subsubsection{Magnetic and thermodynamik properties (poly, Sn flux)}

Some magnetic properties of Sn-flux-grown EuZn$_2$As$_2$ can be found in Tab.~\ref{Tabellemagnet_Zn}.
EuZn$_2$As$_2$ orders antiferromagnetically at $T_{\rm N}=19.6\,\rm K$ with A-type AFM order determined by neutron diffraction at $1.5\,\rm K$ but here it was not possible to distinguish between an inplane or out-of-plane spin orientation \cite{Wang2022}. 
In a further neutron diffraction study at $4\,\rm K$, the A-type AFM structure was confirmed and it was possible to fit the data assuming a spin orientation entirely in the $a-a$ plane \cite{Blawat2022}. 
This finding is in contrast to a later study \cite{Bukowski2022}, where a ferromagnetic contribution to the magnetization was found and M\"ossbauer spectroscopy yielded a canted AFM structure with a canting angle of $58^{\circ}$ towards the $a-a$ plane away from the $c$ axis.
This is again contradictious to a later study \cite{Yi2023}, where the authors claim the absence of a canting based on neutron diffraction but also report on a short range FM order resulting in a locally canted AFM structure in certain domains in a wide temperature region.
Obviously, the spin orientation is quite sensitive to application of magnetic fields since the angle dependence of the magnetoresistance MR$_{ab}$ investigated in  \cite{Blawat2022} showed that the spins can be reoriented in field to an angle of $\approx 45^{\circ}$ between the $a-a$ plane and the $c$ axis.
Nevertheless, the statements on the spin orientation in the ground state in EuZn$_2$As$_2$ remain contradictory and should be subject to detailed studies in the future. 
Specific heat data are shown in \cite{Bukowski2022} and a \neel temperature of $19.1\,\rm K$ is observed. The suppression of the anomaly  in field is consistent with AFM order. At higher fields ($\mu_0H> 2\,\rm T$ ($H\perp c$), $\mu_0H> 3\,\rm T$ ($H\parallel c$)), the position of the broad peak in $C_{\rm mag}(T)$ shifts to higher $T$ which is a typical behaviour of FM materials.
Furthermore in \cite{Bukowski2022}, a large magnetocaloric effect is observed which is comparable to that in e.g. EuO \cite{Ahn2005}.

\subsubsection{Transport and electronic properties (poly, Sn flux)}

At ambient pressure, a negative MR is observed after suppressing the resistivity peak at $T_{\rm N}$ with increasing field \cite{Wang2022}. It is found that, compared to EuCd$_2$As$_2$, the nonlinear anomalous Hall effect is smaller in EuZn$_2$As$_2$ \cite{Wang2022}. 
The material is insulating up to a pressure of $\approx 3\,\rm GPa$ \cite{Luo2023}.
The authors of \cite{Luo2023} find that the resistivity of their samples is higher than the one reported previously \cite{Wang2022} and assign this difference to a possibly stronger hole doping in the latter case.
Furthermore, in pressurized EuZn$_2$As$_2$ an insulator-metal transition as well as a topological phase transition is observed \cite{Luo2023}.
Calculations reveal that the insulating state of EuZn$_2$As$_2$ is topologically trivial, while the metallic state exhibits Weyl fermions and Fermi arcs making the pressure- or field-induced insulator-metal transition a topological phase transition \cite{Luo2023}, which is similar to the behaviour found in the case of EuCd$_2$As$_2$. At 1.6\,GPa, the magnetoresistance calculated according to Eqn.~(\ref{MR1}) reaches a maximum
over $1.4 \times 10^4\%$ at $B = 5\,\rm T$ and $T = 2\,\rm K$ \cite{Luo2023}.
It was reported recently, that besides the long-range A-type antiferromagnetic order, a short-range magnetic order is observed in a wide temperature region in magnetic and electron spin resonance measurements which gives rise to an additional topological Hall contribution \cite{Yi2023}.

\subsubsection{Magnetic properties (salt flux)}

In EuZn$_2$As$_2$ grown from salt flux  ferromagnetic order is observed below $T_{\rm C}=42\,\rm K$ \cite{Chen2024_Zn} and some magnetic properties are shown in Tab.~\ref{Tabellemagnet_Zn} for comparison with the AFM case. In $M(H)$, saturation is reached for $H\perp c$ already at $\approx 0.1\,\rm T$ and for $H\parallel c$ at $\approx 2.5\,\rm T$. As the few single crystals grown so far were small, brittle and fragile \cite{Chen2024_Zn} no further studies of thermodynamic or transport properties were carried out so far.

\begin{figure*}[htbp]
    \centering
    \includegraphics[width=1.0\textwidth]{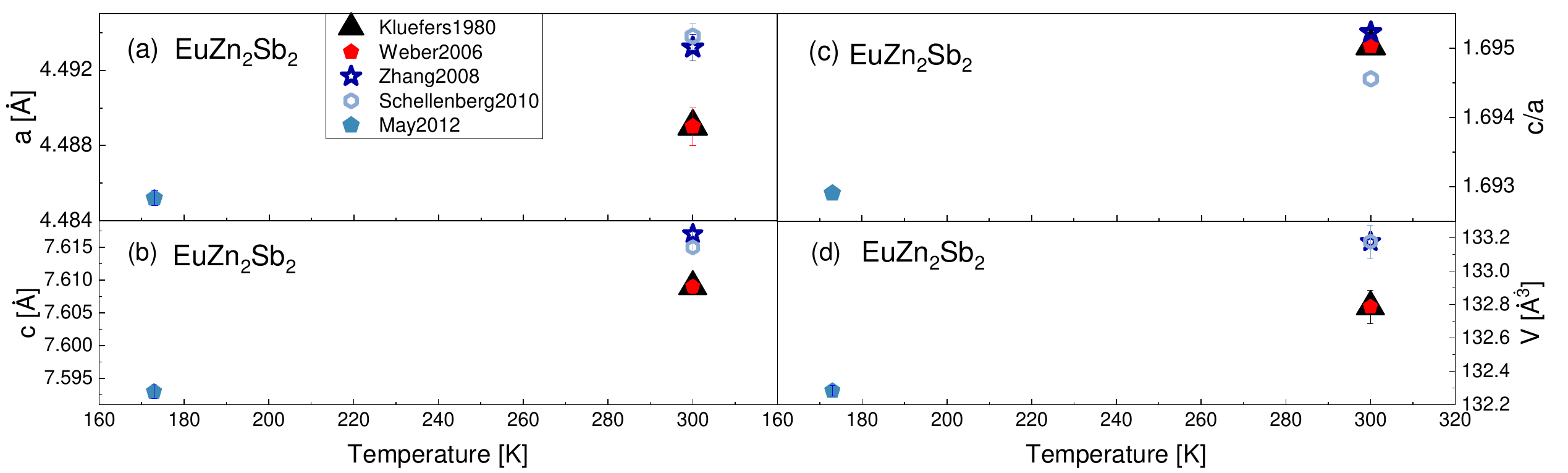}
    \caption{EuZn$_2$Sb$_2$: (a,b) Lattice parameters of poly crystalline samples prepared in Al$_2$O$_3$ (triangles) \cite{Kluefers1980} as well as in Ta (open hexagons) \cite{Schellenberg2010} or in a carbon crucible (asterisk) \cite{Zhang2008}. The lattice parameters of single crystals grown by Bridgman method (Zn-Sb flux) \cite{Weber2005,Weber2006,May2012} are indicated by closed pentagons. (c) $c/a$ ratio and (d) the volume $V$ of the unit cell of the same samples.  }
    \label{fig:4}
\end{figure*}

\subsection{EuZn$_2$Sb$_2$}
\subsubsection{Crystal growth, structural and chemical   characterization}
EuZn$_2$Sb$_2$ was prepared in poly crystalline form in Al$_2$O$_3$  \cite{Kluefers1980}, C \cite{Zhang2008, Zhang2010} or Ta crucibles \cite{Schellenberg2010} and single crystals were grown stoichiometrically by Bridgman method \cite{Weber2005, Weber2006}, from ZnSb \cite{May2012} or Sb \cite{Singh2024} flux using Al$_2$O$_3$ crucibles.
It has to be noted, that samples fron early studies contained a significant amount of a side phase (30-50\% ZnSb) \cite{Weber2006}.
The $a$ and $c$ parameters of EuZn$_2$Sb$_2$, as well as the  $c/a$ ratio and the volume of the unit cell for samples from different sources are depicted for comparison in Fig.~\ref{fig:4}. 
As it can be seen from the data, the largest unit cell volumes at room temperature are found for poly crystalline samples prepared in C \cite{Zhang2008} or Ta crucibles \cite{Schellenberg2010}. Poly crystalline samples \cite{Kluefers1980} as well as single crystals \cite{Weber2006} grown in Al$_2$O$_3$ crucibles exhibit slightly lower $a$, $c$ parameters and unit cell volumes.  
For samples grown from Sb flux using Al$_2$O$_3$ crucibles which were investigated in \cite{Singh2024} no lattice parameters are given.
According to EDX, the preparation of poly crystalline material in carbon crucible \cite{Zhang2010} revealed a close to ideal chemical composition, while the crystal growth from Sb flux (Al$_2$O$_3$ crucible) yielded samples with a slight Eu excess \cite{Singh2024}.

\subsubsection{Magnetic and thermodynamik properties}

The magnetic properties of AFM  EuZn$_2$Sb$_2$ are listed in Tab.~\ref{Tabellemagnet_Zn}.
Antiferromagnetic order below $T_{\rm N}=13.3\,\rm K$ was reported in \cite{Weber2005}.
From magnetization measurements with fields transverse and perpendicular to the $c$-axis different authors deduce an alignment of the magnetic moments along the $c$-axis \cite{Weber2005, May2012} which is contradictious to a  study, where inplane-moment alignment was reported and a spin-flop transition was observed for $H\perp c$ and $\mu_0H=0.05\,\rm T$ \cite{Weber2006}.
The heat capacity shows a second order phase transition at $T_{\rm N}=13.3\,\rm K$ and no additional phase transition \cite{Weber2005}. 
Heat capacity data can be fitted using a combined Debye and Einstein model according to Eqn.~(\ref{HCFIT}) which yielded $\gamma=43.3\,\rm mJ/molK^2$, $\Theta_D=247.68(7)$\,K, and $\Theta_E=77.31(2)$\,K for EuZn$_2$Sb$_2$ \cite{Singh2024}.

\subsubsection{Transport, thermoelectric and electronic properties}
In early resistivity and Hall measurements a semimetallic state and a low carrier concentration $n_{cc} \approx
10^{19}\,\rm cm^3$ was found \cite{Weber2005}.
The temperature dependence of the resistivity was studied in \cite{May2012, Singh2024} with similar results.
Recently, an unconventional anomalous Hall effect was observed in the magnetically ordered  state of EuZn$_2$Sb$_2$ \cite{Singh2024}. Despite the collinear magnetic structure of the material, a scaling of unconventional Hall conductivity with the longitudinal conductivity, and the magnitude of the Hall angle was found indicating spin chirality within domain walls which the authors propose to be the cause for the unconventional anomalous Hall effect.  
The authors point out that maximum in the AHE is strongly sample dependent \cite{Singh2024}.
The thermoelectric properties were studied in \cite{Zhang2008} and the material was characterized as a p-type semiconductor considering the Seebeck and Hall coefficients.
A carrier concentration of $n=2.77\cdot 10^{19}\rm cm^{-3}$ at $300\,\rm K$ was determined from the Hall coefficient $R_H=1/ne$, where $e$ is the charge of an electron.
A high power factor ($\alpha^2\sigma$ with Seebeck coefficient $\alpha$ and $\sigma$ being the electrical conductivity) and a low thermal conductivity $\kappa$ result in a high thermoelectric figure of merit $ZT=\alpha^2\sigma T/\kappa$ which increases from $ZT(300\,\rm K)=0.207$ to $ZT(700\,\rm K)=0.919$ \cite{Zhang2008}.
The figure of merit can be optimized by substitution at the Eu \cite{Takagiwa2017} or Zn sites: Eu(Zn$_{0.9}$Cd$_{0.1}$)$_2$Sb$_2$ shows a maximum $ZT$ of 1.06 at 650\,K \cite{Zhang2010}.

\subsection{EuCd$_2$P$_2$}
\subsubsection{Crystal growth, structural and chemical   characterization}
EuCd$_2$P$_2$ was prepared in poly crystalline form in Al$_2$O$_3$ or Ta crucibles \cite{Artmann1996, Schellenberg2011} and single crystals were grown from tin flux using Al$_2$O$_3$ or C crucibles  \cite{Wang2021, Usachov2024, Chen2024_Cd} or from salt flux in SiO$_2$ \cite{Chen2024_Cd}.
The $a$ and $c$ parameters of EuCd$_2$P$_2$, as well as the  $c/a$ ratio and the volume of the unit cell for samples from different sources are depicted for comparison in Fig.~\ref{fig:5}. 

It is obvious that poly crystalline samples (triangles) \cite{Artmann1996, Schellenberg2011}, which were prepared without usage of a flux show slightly larger unit cell volumes when compared to Sn-flux-grown single crystals (closed circles) \cite{Wang2021, Usachov2024, Chen2024_Cd}. Comparing the poly crystalline samples among each other, the lattice parameter $c$ is higher for the sample which was prepared in a Ta crucible \cite{Schellenberg2011}.
From the temperature dependent data of Sn-flux-grown samples \cite{Usachov2024}, we extrapolated the temperature dependence of the lattice parameters of Sn-flux-grown samples down to lower temperatures assuming linear behaviour down to $\approx 120\,\rm K$ (dashed lines in Fig.~\ref{fig:5}).
We find that the $a$ lattice parameter at 150~K of single crystals grown from salt flux (closed squares) \cite{Chen2024_Cd} is 0.05\% smaller than expected while the $c$ parameter matches the expected value estimated from the temperature dependent data of Sn-flux-grown samples \cite{Usachov2024}. 
Therefore, the volume of the unit cell in case of salt flux usage is $\approx 0.3\%$ smaller (at 150~K) when compared to the expected value at the same temperature in case of Sn flux usage. 
The analysis by single crystal x-ray diffraction reveals that the samples which were studied by \cite{Chen2024_Cd} exhibit a tiny amount of $\approx 0.2\%$ Eu vacancies.

\subsubsection{Magnetic and thermodynamic properties (poly, Sn flux)}

EuCd$_2$P$_2$ grown from Sn flux orders AFM below $T_{\rm N}\approx 11\,\rm K$ in an A-type AFM structure with moments in the $a-a$ plane \cite{Wang2021, Sunko2023, Usachov2024}. For an overview about some magnetic properties of the material, see Tab.~\ref{Tabellemagnet}.
A combination of sensitive probes of magnetism (resonant x-ray scattering and magneto-optical polarimetry) revealed that ferromagnetic order onsets
above $T_{\rm N}$ in the temperature range of the resistivity peak \cite{Sunko2023}. This was further supported through observation of a reorientation of magnetic domains at temperatures up to $\approx 2\,T_{\rm N}$ which confirms the presence of magnetic order above $T_{\rm N}$ \cite{Usachov2024}.
The field dependence of the heat capacity shows a shift of the heat-capacity peak towards higher temperatures which is correlated to the shift of the maximum in the resistivity. It was observed that the shift is significantly larger than that predicted by a mean-field model for an AFM material which indicates an increase of ferromagnetic interactions
above $T_{\rm N}$ \cite{Usachov2024}.

\subsubsection{Transport and electronic properties (poly, Sn flux)}

EuCd$_2$P$_2$ grown from Sn flux shows a peak in the resistivity above the \neel temperature and a CMR effect \cite{Wang2021, Sunko2023, Zhang2023,  Chen2024_Cd, Usachov2024}. Remarkably, the temperature $T^{\rho}$ at which the peak occurs obviously depends on the sample preparation since samples from different sources show different peak temperatures ($T^{\rho}=18$\,K \cite{Wang2021, Sunko2023, Chen2024_Cd}, $T^{\rho}=14$\,K \cite{Usachov2024, Zhang2023}).
The magnetoresistance, calculated according to Eqn.~(\ref{MR1}), is larger than $-10^{-3}\%$ in less than 1\,T \cite{Wang2021} or calculated according to Eqn.~(\ref{MR2}), it is $-93\%$ at 1\,T \cite{Usachov2024}.
For antiferromagnetic EuCd$_2$P$_2$ grown from Sn flux, the hole density was calculated from the Hall resistivity, $\rho_{xy}$, to be $\approx 3.6\times 10^{18}\rm cm^{-3}$ at 150\,K (corresponding to 0.0002 Eu vacancy/f.u.). This is only a tenth of the value of salt-flux grown FM-EuCd$_2$P$_2$ \cite{Chen2024_Cd}.
An inverted hysteresis was observed in the region of the resistivity peak which was interpreted as promotion of ferromagnetism by coupling of localized spins and itinerant carriers and the resulting localization of carriers was confirmed by optical conductivity measurements \cite{Sunko2023}. A further optical study \cite{Homes2023} gave hints that that the resistivity maximum and subsequent carrier localization is due to the formation of ferromagnetic domains below $2\,T_{\rm N}$ that result in spin-polarized clusters due to spin-carrier coupling.
The band structure of the material was investigated using angle resolved photoemission spectroscopy (ARPES) and revealed metallic behaviour \cite{Zhang2023} which is in contrast to a later study where no metallicity was found \cite{Usachov2024}.

\begin{figure*}[htbp]
    \centering
    \includegraphics[width=1.0\textwidth]{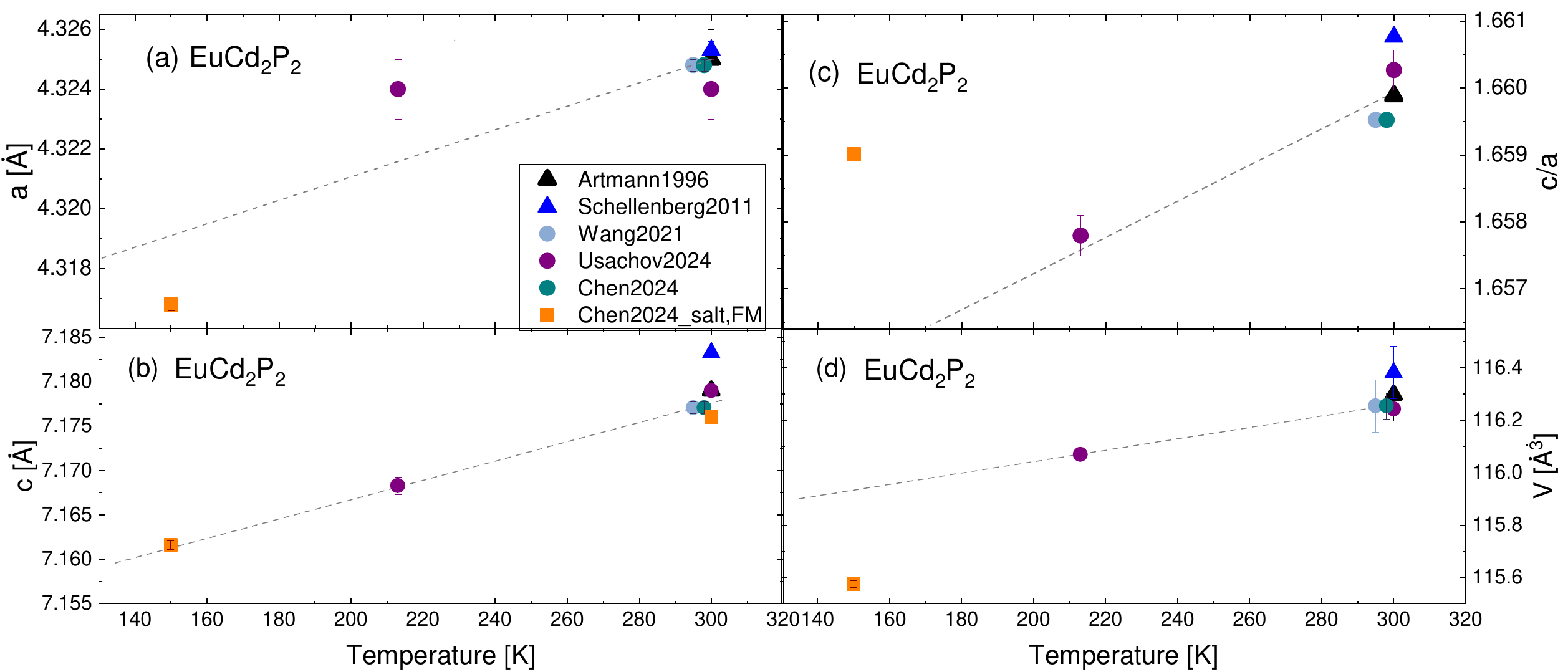}
    \caption{EuCd$_2$P$_2$ (a, b) Lattice parameters $a, c$ from different sources where triangles indicate data determined from measurements on poly crystalline samples \cite{Artmann1996, Schellenberg2011} and circles those from single crystals grown out of  Sn flux \cite{Wang2021, Usachov2024, Chen2024_Cd}. The orange squares indicate the values measured on samples grown from salt flux \cite{Chen2024_Cd}. (c) $c/a$ ratio and (d) the volume $V$ of the unit cell of the same samples. Assuming linear changes down to 100~K, dashed lines mark the extrapolated decrease of the lattice parameters, the $c/a$ ratio and the volume of the unit cell.}  
    \label{fig:5}
\end{figure*}

\subsubsection{Magnetic, transport and electronic properties (salt flux)}

In EuCd$_2$P$_2$ grown from salt flux  ferromagnetic order is observed below $T_{\rm C}=47\,\rm K$ \cite{Chen2024_Zn}. Some magnetic properties are shown in Tab.~\ref{Tabellemagnet} for comparison with the AFM case.
The magnetoresistance of FM-EuCd$_2$P$_2$ is negative below 150\,K and, calculated according to 
Eqn.~(\ref{MR1}), reaches a maximum of MR$=-110\%$ at 30\,K under a field of 9\,T 
\cite{Chen2024_Cd}.
The Hall resistivity $\rho_{xy}(H)$, Eqn.~(\ref{HallRes}) of FM-EuCd$_2$P$_2$, becomes non-linear below $\approx 150 \,\rm K$ due to the increase of a contribution from the anomalous Hall effect and from $\rho_{xy}$ a carrier concentration of $\approx 4.6 \times 10^{19}\rm cm^{-3}$ (corresponding to 0.0026 Eu vacancy/f.u.) at 150\, K was estimated \cite{Chen2024_Cd}.

\subsection{EuCd$_2$As$_2$}

\subsubsection{Crystal growth, structural and chemical characterization}
EuCd$_2$As$_2$ attracts significant attention since it was claimed to be a magnetic Weyl semimetal in the FM polarized state \cite{Rahn2018, Soh2019, Wang2019, Ma2019}. Due to alterations in the magnetic ground state and charge transport that were observed in different studies it soon became clear that application of a small pressure or modifying the crystal growth conditions can change the samples properties strongly \cite{Jo2020, Gati2021, Du2022, Roychowdhury2023} a circumstance which makes it also quite difficult to compare published  results obtained on different samples obtained through different synthesis procedures. 

\begin{figure*}[htbp]
    \centering
\includegraphics[width=1.0\textwidth]{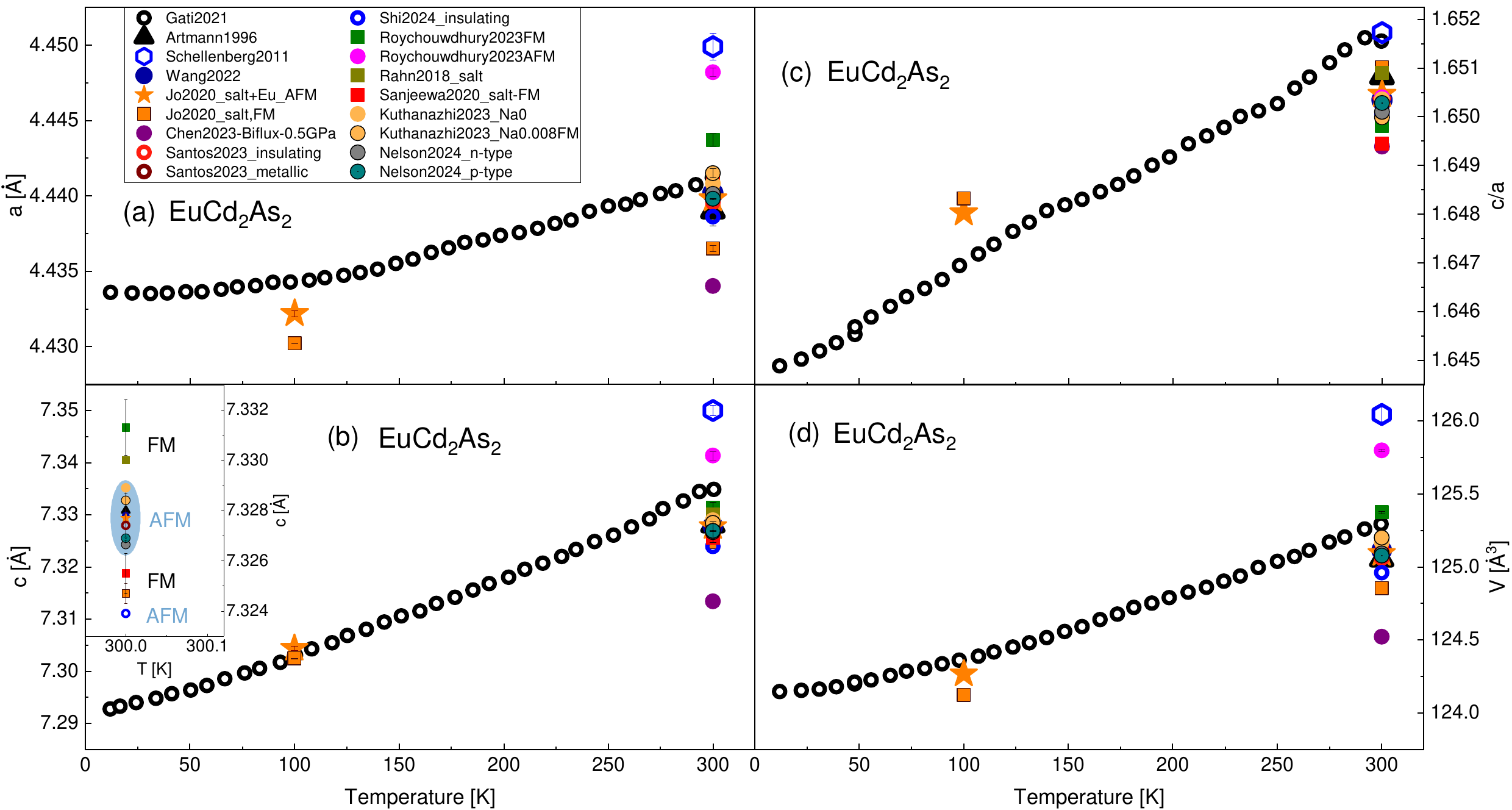}
    \caption{EuCd$_2$As$_2$: Temperature dependence of the lattice parameters of poly crystalline samples prepared in Al$_2$O$_3$ (triangles) \cite{Artmann1996} as well as in Ta (open hexagons) \cite{Schellenberg2011}. The lattice parameters of single crystals grown from Sn flux \cite{Wang2022} are indicated by circles. Orange symbols indicate data determined on samples grown from salt flux without (squares) or with Eu excess (asterisks). (c) $c/a$ ratio and (d) the volume $V$ of the unit cell of the same samples. The inset in (c) shows an enlarged view around the $c$ parameter for Sn-flux-grown samples and the respective magnetic order. }
    \label{fig:6}
\end{figure*}

\begin{figure*}[htbp]
    \centering
\includegraphics[width=0.9\textwidth]{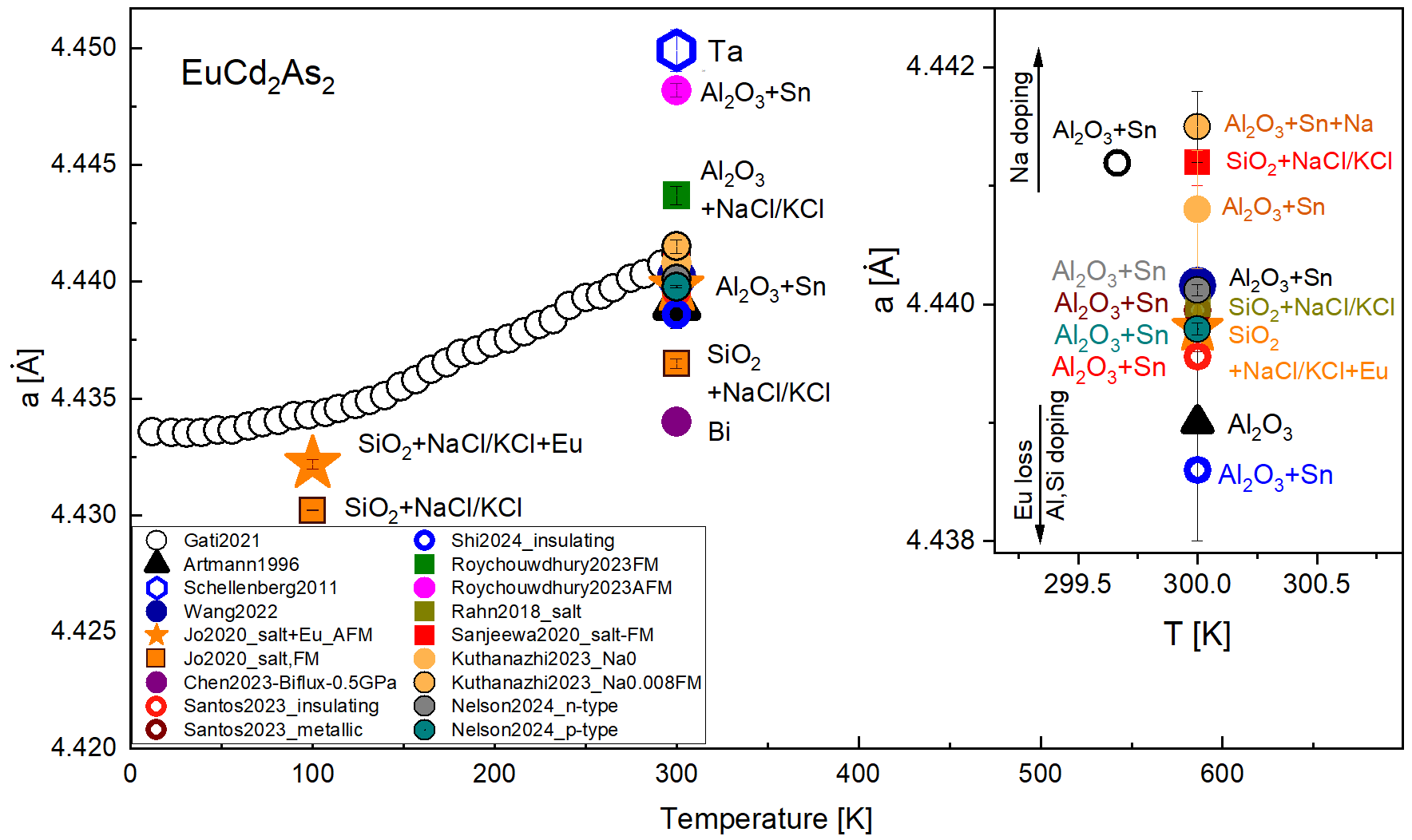}
    \caption{EuCd$_2$As$_2$: $a$ parameter and growth conditions  for samples from different sources (data of the $a$ parameters from Fig.~\ref{fig:6}). The inset shows an enlarged view close to the $a$ parameter of Sn-flux grown samples. Arrows indicate the proposed direction of the shift of the lattice parameter due to chemical pressure.}
    \label{fig:5a}
\end{figure*}

So far, eight different experimental setups were described to prepare EuCd$_2$As$_2$: (i) poly crystalline samples without usage of flux in sealed Ta crucibles \cite{Schellenberg2011} (ii) poly crystalline samples without usage of flux in Al$_2$O$_3$ crucibles \cite{Artmann1996} (iii) 
growth of single crystals from Sn flux in Al$_2$O$_3$ crucibles  \cite{Ma2019, Ma2020, Jo2020, Gati2021, Sun2022, Du2022, Cao2022, Santos2023, Roychowdhury2023, Shi2024} (iv) upon variation of element:Sn-flux ratio \cite{Shi2024} (v) from  Bi flux \cite{Chen2023_CdAs} (vi) from salt flux in SiO$_2$ ampules \cite{Wang2016, Rahn2018, Jo2020, Sanjeewa2020} (vii) from salt flux in Al$_2$O$_3$ crucibles  \cite{Roychowdhury2023} (viii) from salt flux in SiO$_2$ ampules using Eu excess \cite{Jo2020}.
The structural parameters of samples obtained through the different methods of preparation are compared in Figs.~\ref{fig:6} and \ref{fig:5a} from which the following observations can be made and some conclusions can be drawn:

Poly crystalline samples prepared in sealed Ta crucibles \cite{Schellenberg2011} show the largest lattice parameters also larger than the poly crystalline samples prepared in Al$_2$O$_3$ \cite{Artmann1996}. Since Ta is a high-melting element and supposedly inert towards the other elements used in the process at the given "low" synthesis temperature of 822$^{\circ}$C, the losses or doping  during preparation should be lowest compared to all other setups. 

The lattice parameters of samples grown from Sn flux in Al$_2$O$_3$ are quite similar to each other \cite{Ma2019, Ma2020, Jo2020, Gati2021, Sun2022, Du2022, Cao2022, Santos2023, Roychowdhury2023, Shi2024}. Furthermore, they are also similar to the lattice parameters of poly crystalline material when prepared in Al$_2$O$_3$ crucible \cite{Artmann1996}.
During preparation of similar Eu compounds from the elements using Sn flux in Al$_2$O$_3$ crucibles it was observed, that Eu was partially oxidized \cite{Krebber2023, Usachov2024} and that the melt attacks the crucible in self-flux growth \cite{Kliemt2022a}. 
Upon contact of the melt with Al$_2$O$_3$, oxygen is most likely released to the melt which leads to oxidation of Eu which then reduces the amount of Eu that is available for the crystal growth. Furthermore, Al might enter the melt and may cause an unintended doping of the crystals.  

A difference in lattice parameters between annealed poly crystalline material and single crystals grown from flux is  observed for many compounds and sometimes ascribed to a homogeneity range of the material. So far, in the Eu$T_2Pn_2$ ($T=Zn,Cd$; $Pn=P,As$) there is no indication for the existence of such a range.

In \cite{Santos2023}, Eu of varying purity was used to grow samples from Sn flux using Al$_2$O$_3$ crucibles. The usage of higher purity europium (4N) lead to insulating samples with a lower carrier concentration than in the growth where europium with lower purity (2N) was used and which resulted in metallic samples. The lattice parameters of both types of samples are very similar and also similar to other Sn-flux grown samples.  

While AFM crystals grown from Sn flux in Al$_2$O$_3$ crucibles exhibit quite similar lattice parameters, this is not the case for the FM variants grown from salt flux in SiO$_2$ \cite{Jo2020, Sanjeewa2020} or Al$_2$O$_3$ \cite{Roychowdhury2023} as shown in the inset of Figs.~\ref{fig:6}b) and \ref{fig:5a}.  

FM single crystals ($T_C=26.4\,\rm K$) grown from salt flux in SiO$_2$ using a stoichiometric ratio of the elements by Jo {\it et al.} \cite{Jo2020} show an Eu deficiency deduced from magnetization and PXRD data, and a by $0.1\%$ smaller $a$ lattice parameter when compared to Sn-flux-grown samples. Remarkably, the $a$ parameter can be increased to the Sn-flux-grown value again by adding an Eu excess in the initial melt \cite{Jo2020}. This growth procedure (SiO$_2$ ampule, salt flux, Eu excess) also recovers the AFM state ($T_N=9.2\,\rm K$).
These results support the conclusion that usage of Al$_2$O$_3$ crucibles leads to a loss of Eu due to oxidation, meaning that less Eu is available for the growth.

FM single crystals ($T_C=29\,\rm K$) have also been grown from salt flux in SiO$_2$ ampules by Sanjeewa {\it et al.} \cite{Sanjeewa2020}. These authors cannot confirm Eu vacancies in their samples. Interestingly, the $a$ lattice parameter reported in this work for the FM samples is comparable to that of the Sn-flux grown AFM samples. 

An alternative setup was chosen by  Roychowdhury {\it et al.} who used Al$_3$O$_3$ crucibles in combination with salt flux \cite{Roychowdhury2023}. The FM samples ($T_C\approx26\,\rm K$) that resulted from this exhibit lattice parameters closer to that reported by \cite{Schellenberg2011} for poly crystalline samples. In this setup, no direct contact was made with the salt and the SiO$_2$ ampule which prohibited possible Si doping of the melt. Given the fact, that the structural parameters are closer to the "ideal" poly crystalline case, it is also likely that Eu is protected by the salt from oxidation due to the Al$_2$O$_3$ usage and also no/little Al doping occurred. Nevertheless, the authors mention, that $\approx$1\% Eu vacancies would be consistent with their SC-XRD data.
On the other hand a high carrier concentration of 10$^{20}\rm cm^{-3}$  is reported \cite{Roychowdhury2023}, the origin of which might be possible doping of the crystals by Na or K. 

In \cite{Kuthanazhi2023} it is shown that a net FM moment is stabilized by both, Ag and Na doping. The growth was done from Sn flux using highest purity elements like Eu (4N+) in Al$_2$O$_3$ crucibles. The authors assume that Na occupies the Eu sites and found that in Eu$_{1-y}$Na$_y$Cd$_2$As$_2$ ($T_C\approx 17\,\rm K$) the lattice parameter $a$ increases by 0.02\% for a Na doping of $y=0.008$ while the $c$ parameter stays nearly unchanged.
This trend trend is completely consistent with the enhanced lattice parameters reported by \cite{Roychowdhury2023} for FM samples in case of NaCl/KCl flux usage and hints also in case of the samples studied in \cite{Roychowdhury2023} to a small doping of EuCd$_2$As$_2$ by Na or K (or both?) at the Eu sites.

Interestingly, in a high-pressure study \cite{Chen2023_CdAs}, AFM intrinsically insulating EuCd$_2$As$_2$ grown from Bi flux was investigated. The room-temperature lattice parameters are the smallest that are reported so far but it remains unclear whether the small pressure of 0.5~GPa or the growth conditions are responsible for it.
The carrier concentration at 5~K estimated in this work for the insulating samples was 1.7$\times$10$^{17}\rm cm^{-3}$.

In a recent work \cite{Shi2024}, samples were grown in Al$_2$O$_3$ crucibles from Sn flux using different element-to-Sn ratios (1:2:2:10 to 1:2:2:25) and samples with different transport behaviours were obtained. The authors found that the residual carrier density of the samples is highly sensitive to the ratio of Sn flux in the starting materials. The carrier concentration obtained in this way in the insulating samples was $10^{15}-10^{16}\rm cm^{-3}$. As the reported changes are not systematic, most likely the different ratios offer randomly more or less protection for Eu from oxidation depending on how well Eu is protected from getting in contact with Al$_2$O$_3$. 

La doping (highest purity elements usage, Sn flux, Al$_2$O$_3$ crucibles, incorporation of La $<1\%$) was done by Nelson {\it et al.} to produce n-type samples which yielded at most a minor increase of the $a$ parameter when compared to that of their p-type samples \cite{Nelson2024}. 



\subsubsection{Magnetic and thermodynamic properties (poly, Sn flux)}

Sn-flux-grown EuCd$_2$As$_2$ shows AFM order of the Eu$^{2+}$ ions below $T_{\rm N}= 9.5\,\rm K$ \cite{Schellenberg2011}, see Tab.~\ref{Tabellemagnet} for an overview about some magnetic properties reported in different studies. Initially, FM order was detected at 16\,K \cite{Artmann1996} which was later ruled out in a M\"ossbauer study of single crystals and attributed to Eu-containing impurities due to oxidation of the poly crystalline sample \cite{Schellenberg2011}.  

The material was identified as a low carrier density semimetal \cite{Wang2016} and attracted the interest of the community \cite{Rahn2018} as being related to Cd$_3$As$_2$ which is the first example of a bulk Dirac semimetal \cite{Liu2014}. In 2016 through REXS experiments and {\it ab initio} electronic structure calculations, EuCd$_2$As$_2$ was identified as a candidate Dirac semimetal with an A-type AFM structure with magnetic moments oriented in the $a-a$ plane and large magnetoresistive effects associated with field-induced changes in the magnetic structure and domain distributions \cite{Rahn2018}. 
For some time it was under debate whether in the ground state the magnetic moments are aligned along $c$ \cite{Wang2016} or along an in-plane direction where the C3 symmetry would be broken \cite{Rahn2018}. 
Later, Cao {\it et al.} discussed a possible spin canting of 30$^{\circ}$ away from the $c$ axis which would be consistent with an observed maximum of the peak of the Hall resistivity \cite{Cao2022}. 
Since the Weyl semimetal state \cite{Ma2019} was thought to be accessible in EuCd$_2$As$_2$ in a ferromagnetic structure with moments aligned along the $c$ axis \cite{Rahn2018, Soh2019}, its magnetic structure as well as options to change it were further investigated. Experimental evidence for a field-induced metamagnetic transition was found \cite{Sun2022} and it turned out that a small amount of Ba ($\approx 3\%–10\%$) substitution leads to a small out-of-plane canting of the Eu moment \cite{Sanjeewa2020}.

Soon later it became clear, that a FM variant of EuCd$_2$As$_2$ can be grown when using different synthesis conditions \cite{Jo2020} which shows that the FM state is very close to the AFM one and demonstrates a high tunability of the magnetic properties of the material. The authors assigned this change of the magnetic ground state to a presence of an Eu deficit in the FM samples \cite{Jo2020}.
Furthermore, it was found that a magnetic state with a net ferromagnetic moment can be stabilized by substitution with both Ag and
Na. Therefore, chemical substitution and the subsequent changes in band filling can be used to tune the magnetic ground state and to stabilize a ferromagnetic phase in EuCd$_2$As$_2$ \cite{Kuthanazhi2023}.

In 2023, Santos-Cottin {\it et al.} performed the crystal growth under extremely pure conditions using extra pure starting materials like Eu 4N to investigate the influence of impurities on the transport behaviour. This yielded insulating (growth under extra pure conditions) as well as metallic (growth under standard conditions) samples. The author showed that while the transport properties change, the magnetic transition temperature stays at $T_{\rm N}=9.5\,\rm K$ in the insulating case and that both samples show the same magnetization behaviour \cite{Santos2023}.

\subsubsection{Transport and electronic properties (poly, Sn flux)}

In 2019, using $\mu SR$, magnetization, transport and ARPES measurements, a Weyl semimetal state was identified in the paramagnetic phase of EuCd$_2$As$_2$ whose presence was attributed to quasistatic and quasi-long-range ferromagnetic fluctuations \cite{Ma2019}. 

Based on DFT calculations \cite{Wang2019}, it was expected that AFM-EuCd$_2$As$_2$ is a topological insulator while two pairs of Weyl points exist in FM-EuCd$_2$As$_2$.
From further DFT+U calculations it was deduced, that EuCd$_2$As$_2$ is a Dirac semimetal if it is in the A-type AFM phase with out-of-plane magnetic moments. For both configurations (out-of-plane as well as in-plane moments), the band structure is very similar, but for the in-plane AFM order, a gap of $\approx 1\,\rm meV$ opens at the bulk Dirac cone point and the system would evolve into an AFM topological insulator \cite{Ma2020}. 
After the prediction of the presence of a single pair of Weyl points in case of a FM Eu-spin configuration along the $c$ axis \cite{Ma2020}, the system was tuned by applying hydrostatic pressure \cite{Gati2021}. In this study by Gati {\it et al.} it was found that the magnetic moments are aligned in the $a-a$ plane in the ground state (AFM$_{ab}$) and that a pressure of $\approx$ 2\,GPa changes the alignment to a FM one with moments inplane (FM$_{ab}$) \cite{Gati2021}. This is consistent with the results presented later in \cite{Yu2023}. Furthermore, it was predicted, that a FM$_c$ alignment can be achieved by application of pressure of $\approx 23\,\rm GPa$ if the system stays magnetic \cite{Gati2021}.
A further study in 2022 \cite{Du2022} of the resistivity of EuCd$_2$As$_2$ single crystals under hydrostatic pressure, yielded an insulating dome in a small pressure range ($\approx 1.0$\,GPa- 2.0\,GPa), while otherwise metallic transport is observed. It was found that the insulating state can be suppressed by a magnetic field of 0.2 T for $H \perp c$ which leads to a colossal negative magnetoresistance (MR) of $\approx 10^5\%$ at 0.3\,K. Based on the results of first principles calculations, the authors find that these experimental observations may arise
from two topological phase transitions: (i) at $\approx 1\,$GPa a transition from an AFM topological insulator to an AFM trivial insulator (TrI) and (ii) at $\approx 2\,$GPa a transition from an AFM TrI to a FM Weyl semimetal (WSM) \cite{Du2022}. The transport properties were further investigated by \cite{Cao2022} who have observed a giant nonlinear intrinsic
AHE in EuCd$_2$As$_2$. 

Until 2023, EuCd$_2$As$_2$ was widely accepted as a topological semimetal in which a Weyl phase can be induced by application of an external magnetic field.
This view was challenged by Santos-Cottin {\it et al.} \cite{Santos2023} who studied samples of different purity of the material by a combination of electronic transport, optical spectroscopy, and excited-state photoemission spectroscopy. While samples grown under standard-purity conditions using Eu (2N) are metallic, samples grown using high-purity Eu (4N) are insulating which lead to the conclusion that EuCd$_2$As$_2$ is not a topological semimetal, but rather a semiconductor with a band gap of 770 meV  \cite{Santos2023}. 

In a high-pressure study, AFM intrinsically insulating EuCd$_2$As$_2$ grown from Bi flux was investigted and a structural phase transition from trigonal $P\overline{3}m1$ to monoclinic $C2/m$ was detected above 20\,GPa. The analysis of the structural parameters revealed that this transition is accompagnied by the emergence of a FM metallic state with $T_C\approx 150\,\rm K$ beyond $\approx 24\,$GPa \cite{Chen2023_CdAs}. The transition from an AFM state with moments inplane to a FM state with moments inplane at $\approx 2\,\rm GPa$ found for metallic samples \cite{Gati2021} was not confirmed in this study with insulating samples being investigated \cite{Chen2023_CdAs}. The high-pressure studies \cite{Chen2023_CdAs, Yu2023} consistently show, that the transition temperature increases upon application of pressure.

The discussion about the nature of EuCd$_2$As$_2$ now hinges around the question of whether there is a band overlap between the conduction band minimum and the valence band maximum, which enables a Weyl semimetal state with ferromagnetic spin alignment, or whether the material is rather a trivial semiconductor as proposed by  \cite{Santos2023}. A recent study of samples with a low carrier density of $2\times 10^{15}{\rm cm}^{-3}$ revealed a carrier density dependence of the anomalous Hall effect which indicates the absence of Weyl nodes in the material \cite{Shi2024}.
Furthermore, p-type as well as n-type carrier densities of 6.0$\times 10^{17}$\,cm$^{-3}$ at low temperature and 8.3$\times 10^{17}$\,cm$^{-3}$ at high temperature were determined on the samples investigated by Nelson {\it et al.} and the semiconducting band gap in EuCd$_2$As$_2$ was revealed via n-type lanthanum doping \cite{Nelson2024}.

\subsubsection{Magnetic properties (salt flux)}

While AFM EuCd$_2$As$_2$ grown from Sn flux in many different studies, see Tab.~\ref{Tabellemagnet} for comparison,  shows a quite similar \neel temperature in all works, this is not the case for FM salt-flux-grown EuCd$_2$As$_2$ where $T_C$ ranges between $\approx 26\,\rm K$ and $\approx 30\,\rm K$ \cite{Jo2020, Sanjeewa2020, Taddei2022, Roychowdhury2023}.
Also, for the salt-flux samples, different behaviour in field is reported concerning the fields at which saturation occurs in $M(H)$ for the different crystallographic directions. This sample dependency is probably caused by different sorts of doping stemming from the different experimental setups as detailed above. 
Neutron diffraction was performed on a sample $^{153}$Eu$^{116}$Cd$_2$As$_2$ grown from isotopes with a lower absorption (but also lower purity) to get insights into the magnetic structure of the material.
The FM $^{153}$Eu$^{116}$Cd$_2$As$_2$ grown as described in \cite{Sanjeewa2020} with $T_C\approx30\,\rm K$ was investigated and  revealed the best fit to the data for a magnetic space group $C2'/m'$ and confirmed the Eu moments pointing along the [210] direction in plane and canted $\approx 30^{\circ}$ out of plane \cite{Taddei2022}. In comparison, a 60$^{\circ}$ out-of-plane canting was proposed by \cite{Cao2022} for AFM EuCd$_2$As$_2$.

\subsubsection{Thermodynamic, transport and electronic properties (salt flux)}
The transition temperature of $T_C=26.4\,\rm K$ of salt-flux-grown EuCd$_2$As$_2$ was determined by Jo {\it et al.} using electrical transport and heat capacity measurements \cite{Jo2020}. The transverse Hall effect of samples with $T_C=29\,K$ was investigated and an AHE that likely results from an intrinsic contribution was found \cite{Sanjeewa2020}. In a later study of metallic samples with a high carrier density of $\approx 10^{20}$\,cm$^{-3}$ by Roychowdhury {\it et al.}, a maximum negative magnetoresistance (nMR) of $\approx 54\%$, calculated according to Eqn.~\ref{MR2}, at 26~K for both configurations ($H\parallel I$ and $H\perp I$) in the temperature range
of 5~K to 50~K which is commonly observed in FM systems due to the field driven suppression of thermal spin fluctuations.
In case of Weyl materials, an nMR is expected for $H\parallel I$ caused by chiral anomalies \cite{Roychowdhury2023}.
Nevertheless, the authors conclude from their results that the anomalous Hall conductivity and the Nernst effect have an intrinsic contribution resulting from a Berry curvature in the material. Also, supported by DFT calculations, a large topological Hall effect (THE) arises from changes in the band structure caused by spin canting \cite{Roychowdhury2023}.

\subsection{EuCd$_2$Sb$_2$}
\subsubsection{Crystal growth, structural and chemical characterization}

\begin{figure*}[htbp]
\centering
\includegraphics[width=1.0\textwidth]{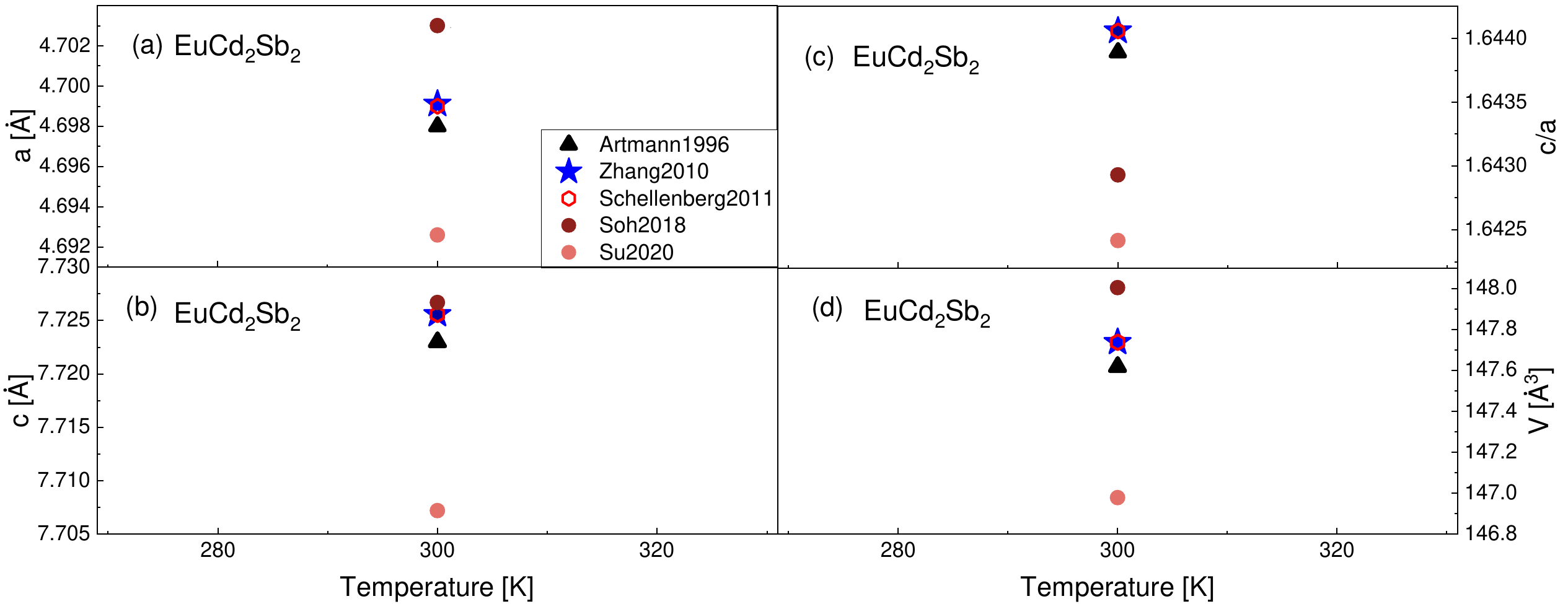}
\caption{EuCd$_2$Sb$_2$: (a,b) Lattice parameters of poly crystalline samples prepared in Al$_2$O$_3$ (triangles) \cite{Artmann1996} as well as in Ta (open hexagons) \cite{Schellenberg2011} and carbon (asterisks) \cite{Zhang2010}. The lattice parameters of single crystals grown from iodine via vapour transport \cite{Soh2018,Su2020} are indicated by circles. (c) $c/a$ ratio and (d) the volume $V$ of the unit cell of the same samples.}
    \label{fig:7}
\end{figure*}

The structural parameters for different kinds of preparation are compared in Fig.~\ref{fig:7} and summarized in Tab.~\ref{Gitterkonstanten2}. 
So far, mainly poly crystalline material has been studied with respect to its thermoelectric properties \cite{Artmann1996, Zhang2010, Zhang2010a, Zhang2010b, Goryunov2012, Schellenberg2011}. 
The $a$ lattice parameters of samples prepared in inert Ta \cite{Schellenberg2011} and carbon \cite{Zhang2010} are slightly larger than that found for samples prepared in Al$_2$O$_3$ \cite{Artmann1996}.
Despite their preparation of poly crystalline material in carbon crucible (which should not cause losses of Eu due to oxidation) it yielded samples with an Eu deficiency as reported by \cite{Zhang2010, Zhang2010b}, but their lattice parameters match that published by \cite{Schellenberg2011} for preparation in inert Ta crucible.
Vapour transport \cite{Soh2018, Su2020} experiments were carried out to grow single crystals using iodine as transport agent and remarkably deviating values of the $a$ parameters were obtained which are maybe caused by losses during prereaction in Al$_2$O$_3$ leading to an Eu deficiency/Al doping (smaller $a$ parameter) or maybe slight doping by larger iodine at the Sb site (larger $a$ parameter).

\subsubsection{Magnetic and thermodynamic properties}
The heat capacity of poly crystalline EuCd$_2$Sb$_2$  in zero field shows a sharp anomaly at $T_{\rm N}=7.13\,\rm K$ \cite{Zhang2010a}, 
for a summary of some magnetic properties see  Tab.~\ref{Tabellemagnet}. 
REXS at the Eu $M_5$ edge reveals an antiferromagnetic structure below $T_N=7.4\,\rm K$ with propagation vector ${\bf k}=(0\,0\, 1/2)$ and spins aligned in the $a-a$ plane \cite{Soh2018}. Poly crystalline trigonal EuCd$_2$Sb$_2$ was prepared from the stoichiometric elements in Al$_2$O$_3$ crucibles and afterwards treated for several hours under a pressure of 6~GPa at $T=600^{\circ}$C. It was found that tetragonal EuCd$_2$Sb$_2$ can be obtained through this procedure and magnetization measurements confirmed AFM ordering with \neel temperature around 16~K \cite{Jimenez2023}.
The low temperature specific heat of tetragonal EuCd$_2$Sb$_2$ yields a Sommerfeld coefficient of $0.33(2)\,\rm J/molK^2$ which shows that the system is a potential heavy-fermion Kondo insulator  \cite{Jimenez2023}.

\subsubsection{Thermoelectric, transport and electronic properties}

EuCd$_2$Sb$_2$ shows a figure of merit of $ZT\approx 0.66$ at 616~K \cite{Zhang2010}, a Hall coefficient of +1.426 cm$^3$/C, a carrier mobility of 182.40 cm$^2$/Vs and a carrier density of $4.38\times10^{18}\rm cm^{-3}$ which characterize the compound as a good thermoelectric material \cite{Zhang2010a}.  
Parallel to the predictions of Weyl physics in EuCd$_2$As$_2$ \cite{Wang2019, Soh2019, Ma2019}, a magnetic exchange induced Weyl state was also found in the case of EuCd$_2$Sb$_2$ \cite{Su2020}. The authors report that an external field $>3.2$~T can align all
spins to be fully polarized along the $c$-axis and also in EuCd$_2$Sb$_2$ a spin polarized state can be reached and according to {\it ab initio} calculations, a phase transition from a small gap AFM topological insulator to a spin polarized WSM may occur \cite{Su2020}.


\begin{table*}[bpth]
\begin{center}
\begin{tabular}{|c|ccccccccccc|}
\hline\hline
Comp. & Flux/&magnetic&$T_M$&CMR&$\Theta_W^{\perp c}$&$\Theta_W^{\parallel c}$ &$\mu_{\rm eff}^{\perp c}$         &   $\mu_{\rm eff}^{\parallel c}$ &moment   	 &  $\mu_0H_{\rm sat}$& Ref.\\
         & method  &order    &    [K] &    [K]   &[K] & [K]        &[$\mu_{B}$]&[$\mu_{B}$]&dir.&[T]&\\
\hline
\hline	
       &Sn         & AFM     &23 &  -     &19.2 &41.9 & 8.61   & 7.74 &$a$ & -& \cite{Berry2022}\\
EuZn$_2$P$_2$& Sn	&A-type$^{\dag}$      &23.7 & + &$33\pm1$ &$33\pm1$&$7.84\pm0.02$&$7.84\pm0.02$  &$a/c$ canted & 1.3 ($\parallel a$),  & \cite{Krebber2023}\\
& & AFM     &&&&&&  & $40^{\circ}\pm 10^{\circ}$&2.5 ($\parallel c$)  & \cite{Krebber2023}\\
& Sn	& AFM    & 23.5&   -&27.45&-&8.19&- & $a$ & -& \cite{Singh2023}\\
& Sn	& AFM  &23.5 &-&27.7&- &8.6 &-& - &  -    &\cite{Chen2024_Zn}\\
& Sn	& AFM   &23.3 &+&$27.8\pm0.2$&$24.8\pm0.3$ &$7.74\pm0.01$ &$7.87\pm0.01$&  $30^{\circ}$$^{\dag\dag}$  & 1.3 ($\parallel a$),&\cite{Rybicki2024}\\
&pressure 	&    && && & &&  &  2.5 ($\parallel c$)    &\cite{Rybicki2024}\\
\hline
EuZn$_2$P$_2$ & salt & FM  &72&- &78.7 &78.7 & 8.2 &8.1   &    $a-a$ &$0.1~ (\perp c)$  & \cite{Chen2024_Zn}\\
& &   & && & & &   &      plane&$2.5~  (\parallel c)$  & \cite{Chen2024_Zn}\\
\hline
\hline
 &  poly  &AFM & 16.5&  -&	-&-& - 	&- &-&-& \cite{Goryunov2012}\\
 &Sn &A-type$^{\ddagger}$  &19.6 & +&20.2  & 17.9 &8.28 &	8.39   &	$\parallel$ or $\perp c$$^{\ddagger}$	& 2.5 ($\perp c$)& \cite{Wang2022}\cite{Du2022}\\
 &&AFM&&&&&&&&3.6 ($\parallel c$)&\cite{Wang2022}\\
EuZn$_2$As$_2$ &Sn&AFM&19 &- &15.1 &17 &8.42&	 8.26  &	$\perp c$$^{\ddagger}$	&-&\cite{Blawat2022}\\
 &Sn&AFM &19.2& - &$23.8\pm0.1$&-&$7.90\pm0.01$&	-  &		canted$^{\dag\dag\dag ,}$$^{\dag\dag}$ & 2.2($\perp c)$&\cite{Bukowski2022}\\
 & && & &&& &	   &	$(58\pm1)^{\circ}$ from $c$  	& 3.4 ($\parallel c$)&\cite{Bukowski2022}\\
&Sn &AFM& 18& -&-&-& -&	7.85   &	-  	& 3.9 ($\parallel c$)&\cite{Luo2023}\\
 &Sn &AFM&$\approx 20$ &-&25& 20 &- &	 -  &	 $\perp c^{\ddagger}$&-&\cite{Yi2023}\\
\hline
EuZn$_2$As$_2$ & salt & FM  &42&- &47.4 &51 &7.5&7.0   & -     &$\approx 0.1~ (\perp c)$  & \cite{Chen2024_Zn}\\
& &   & & & & &&   &      &$\approx 2.5~  (\parallel c)$  & \cite{Chen2024_Zn}\\
\hline
\hline
 &Brid.&AFM&13.3&-&- & -&5.3$^{\diamond}$& 5.3$^{\diamond}$&  $\parallel c$  &$\approx 3.5~  (\parallel c)$  &	\cite{Weber2005} \\
 &&&&&&&&& &$\approx 4.7~  (\perp c)$  &	\cite{Weber2005} \\
 &Brid.&AFM&13.3&-&8.8$^{\diamond}$ & 6.65$^{\diamond}$&5.3$^{\diamond}$& 5.3$^{\diamond}$& $\perp c$  &$\approx 4.4~  (\parallel c)$  &	\cite{Weber2006}\\
 &&&& & &&& &  &$\approx 3.2~  (\perp c)$  &	\cite{Weber2006}\\
EuZn$_2$Sb$_2$ &poly&AFM&13.06&-&7.19(8) &7.19(8) &7.89(4)&7.89(4)& -  &-&	\cite{Zhang2008}  \\
 &ZnSb&AFM&$\approx 13$&-&-7&-7 &7.93(1)&7.93(1)    & $\parallel c$&$\approx 4.75~  (\parallel c)$    &	\cite{May2012} \\
 &&&&& &&&    & &$\approx 3.5~  (\perp c)$    &	\cite{May2012} \\
 &Sb&AFM& 13.3&-&- &11.7 &- & 7.95   & -&$\approx 4.7~  (\parallel c)$&	\cite{Singh2024}\\
  &&& & && & &   & &$\approx 3.5~  (\perp c)$&	\cite{Singh2024}  \\
\hline\hline
\end{tabular}
\end{center}
\caption{\label{Tabellemagnet_Zn} Magnetic properties of trigonal EuZn$_2Pn_2$ compounds, space group $P\overline{3}m1$ (No.164). Standard deviations in brackets. $^{\dag}$from resonant magnetic diffraction, moments canted out of the $a-a$ plane by an angle of about $40^{\circ}\pm 10^{\circ}$ and aligned along the $[100]$ direction.  $^{\dag\dag}$from $^{151}$Eu M\"ossbauer spectroscopy, $^{\ddagger}$from neutron diffraction, $^{\dag\dag\dag}$from magnetization, $^{\textasteriskcentered}$from REXS, $^{\diamond}$corrected due to side phase amount in the sample. }
\end{table*}

 \begin{table*}[bpth]
\begin{center}
\begin{tabular}{|c|ccccccccccc|}
\hline\hline
Comp. & Flux/&magnetic&$T_M$&CMR&$\Theta_W^{\perp c}$&$\Theta_W^{\parallel c}$ &$\mu_{\rm eff}^{\perp c}$         &   $\mu_{\rm eff}^{\parallel c}$ &moment   	 &  $\mu_0H_{\rm sat}$& Ref.\\
         & method  &order    &    [K]  & [K] &[K] & [K]        &[$\mu_{B}$]&[$\mu_{B}$]&dir.&at 2~K [T]&\\
\hline
\hline
  &poly   &FM &30&-& 22.9 & 	22.9   &	7.96& 7.96 	&- &-&\cite{Artmann1996}\\
   &poly   &FM&11.6(5)&-&17.8(5)&17.8(5)& 8.02(1) & 	8.02(1)   &	-&	$\approx 5$&\cite{Schellenberg2011}\\
  &Sn   &AFM&11.3(2)&18& 28 & -  &	8.1(3)&  	-&-&0.16 ($\perp c$)&\cite{Wang2021}\\
   &  &&&  & 	  &	&&  	&&1.6 ($\parallel c$)&\cite{Wang2021}\\
    & Sn &A-type$^{\textasteriskcentered}$&11.5&18& - & 		-  &	-&  	-&-&-&\cite{Sunko2023}\\
    &  &AFM&&  & 	&	  &	&  	&&&\cite{Sunko2023}\\
EuCd$_2$P$_2$  &Sn	&  AFM &10.6&14& 20.89 & 	20.70& 7.93 &	7.92& 	$\perp c$ &0.17 ($\perp c$)& \cite{Usachov2024}\\
&	&  &&  && 	&  &	& 	 &1.04 ($\parallel c$)&\cite{Usachov2024}\\
  &Sn	   &AFM&10.9&14& - & 	-   &	-&  -	& -&-& \cite{Zhang2023}\\
   & Sn   &AFM&11&18& - & 	- &   -&	-  	&-&-&\cite{Chen2024_Cd}\\
\hline
EuCd$_2$P$_2$  &salt& FM   &47 &-& - &48.7&  - &	7.9  	& -& 0.07 ($\perp c$)&\cite{Chen2024_Cd}\cite{Chen2024_Zn}\\
              & &  &         &                       &	      &      &  & &&1.6 ($\parallel c$) &\cite{Chen2024_Cd}\cite{Chen2024_Zn}\\
\hline\hline
  &poly   &FM &16&-& 9.3 & 9.3  &	7.7& 7.7	& -&- &\cite{Artmann1996}\\
  &   &AFM &9.5&-&-  & 	-   &	-&- 	&- & -&\cite{Artmann1996}\\
& poly &AFM&9.5&-&  11.8(5) &11.8(5)	&7.88(1)& 	7.88(1)&-&- &\cite{Schellenberg2011}\\
EuCd$_2$As$_2$ &	Sn &AFM&9.5&&-   & - 	&-   &	-  	& -&-& \cite{Ma2019}\\
 &	Sn &AFM&9.5&-&-   & -& -  &	-  	& -&-&\cite{Ma2020}\\
 &	Sn &AFM &-&-&-   &10.3$\pm 0.05$ &-&7.8$\pm$0.1 &	 $\perp c$ 	 &-&\cite{Jo2020}\\
  &	Sn & AFM&9.3&-& -  &- & -  &	 - 	&$\perp c$ &$\approx 0.7$ ($\perp c$)&\cite{Gati2021}\\
  &	 & &&   & & &  &	  	& &$\approx 1.8$ ($\parallel c$)&\cite{Gati2021}\\
 &	Sn & AFM &9.3&-& - & -& -  &	 -	&$\perp c$&-&\cite{Sun2022}\\
 &	Sn & AFM &9.3&press.& - & -& -  &	 -	&$\perp c$&-&\cite{Du2022}\\
\hline
 &	salt& AFM  &9.2&-&   12.4$\pm 0.2$ &-&7.8$\pm$0.2   	& -&- &$\approx 0.6 (\perp c)$& \cite{Jo2020}\\
 &	 &&&   &	&&   	& & &$\approx 1.8 (\parallel c)$ &\cite{Jo2020}\\
EuCd$_2$As$_2$ &  salt &AFM&9.5&-&  -  &	-&-&  	-& $\parallel c$&- &\cite{Wang2016}\\
 &salt \cite{Schellenberg2011}	&A-type  & -&- &-&  - &-  	& -& $\perp c$ &-&\cite{Rahn2018}\\
 &	  &AFM&   &&  	&   & 	& &  &&\cite{Rahn2018}\\
 \hline
 &	salt& FM   &26.4& -& 27.9$\pm 0.1$ &-&7.6$\pm$0.1   &	-& $\perp c$&$\approx 0.6\,  (\perp c)$&\cite{Jo2020}\\
 &	 &&&   &	&&  	& &&$\approx 1.3\,  (\parallel c)$ &\cite{Jo2020}\\
EuCd$_2$As$_2$&	salt &FM&29&-& 30(3)  &	28(3)&  8.1(8) 	& 8.2(8)& $\perp c$ &0.3 ($\perp c$) &\cite{Sanjeewa2020}  \\
&	 &&&   &&	&   	& &  &1.5 ($\parallel c$) &\cite{Sanjeewa2020}\\
&	salt &FM&$\approx 26$&-& -  &	-&   	-&- &$\perp c$ &  $\approx$ 0.11 ($\perp c$) &\cite{Roychowdhury2023}\\
&	salt &&&&   &	&   	& & & $\approx$ 2.2 ($\parallel c$) &\cite{Roychowdhury2023}\\
\hline
\hline
	 &  poly &FM&12&-&-3.3  & - &	7.37&-&-&  - &\cite{Artmann1996}\\
  &  poly &AFM&7.8&&  &  &	& &&  &\cite{Artmann1996}\\
 &poly&AFM &7.13&- & -3.14(7)    &-&7.83(4)&	-&  	-&3&\cite{Zhang2010a}\\
&  poly  &AFM&  	-&-&-3.14(7)&-& 7.83(4)& 	- &-&- &\cite{Zhang2010b}\\
EuCd$_2$Sb$_2$  &poly &AFM &7.4(5)&-&4.6(5)&-&8.11(1)  & -  &	-& -&\cite{Schellenberg2011}\\
 &  poly  & &7.83&-&  -3&	-&-  	&- &-&- &\cite{Goryunov2012}\\
 & SC   &AFM&  7.4 &	-&-3.8(2)&-& 	8.07(5)&-&-&$\approx 2\, (\perp c)$ &\cite{Soh2018}\\
 & SC \cite{Soh2018}  & AFM&7.5& -  &	-&4.6(5)&-&7.94  	 &-&3.2 $(\parallel c)$&\cite{Su2020}\\
\hline
EuCd$_2$Sb$_2$ &  poly &AFM &16 & -  &	-7.5(5)&-& 8.05(2) 	& -&-&-&\cite{Jimenez2023}\\
$P4/mmm$       &  600$^{\circ}$C&6GPa  &&     &	    	& &             &  &&&\\
\hline\hline
\end{tabular}
\end{center}
\caption{\label{Tabellemagnet} Magnetic properties of trigonal EuCd$_2Pn_2$ compounds, space group $P\overline{3}m1$ (No.164). Standard deviations in brackets. $^{\textasteriskcentered}$from REXS.}
\end{table*}

\section{Discussion of general trends}
AFM as well as FM variants have been grown of  EuCd$_2$P$_2$, EuZn$_2$P$_2$, EuCd$_2$As$_2$ and EuZn$_2$As$_2$ and lattice parameters are published of both variants for the first three of them. Remarkably, the salt-flux-grown FM samples show smaller $a$ parameters ($\approx 0.1\%$) when compared to the Sn-flux grown variants while $c$ is unaffected and matches that of Sn-flux grown samples (orange squares in Figs.~\ref{fig:2}, \ref{fig:5}, \ref{fig:6}). 
Due to this reduction of the $a$ parameter in all salt-growth experiments in SiO$_2$ ampules one might also consider a possible doping by Si at the $Pn$ site. The lattice parameters were compared for the Eu$T_2Pn_2$ compounds and setups in crystal growth experiments and the trends that were identified are now discussed starting with EuCd$_2$As$_2$ as most data are available for this material and focusing on the $a$ parameter which shows the largest variation in salt-growth experiments.

The preparation of {\bf poly crystalline  EuCd$_2$As$_2$} from the stoichiometric elements in inert, sealed Ta crucibles \cite{Schellenberg2011} is considered as the "ideal case" with minimal losses. The $a$ lattice parameter takes a value of $a=4.450\,\rm\AA$ in this case as shown in Fig.~\ref{fig:5a} (open hexagon). 
This "ideal" lattice parameter might change due to positive/negative chemical pressure (doping with smaller/larger atoms) or vacancies. In \cite{Jo2020}, an Eu deficiency in single crystals and a smaller $a$ parameter Fig.~\ref{fig:5a} (orange squares) occur simultaneously. 
The preparation of poly crystalline material from the stoichiometric elements in Al$_2$O$_3$ crucible Fig.~\ref{fig:5a} (black triangle) \cite{Artmann1996} yielded a reduced $a$ lattice parameter when compared to the Ta case \cite{Schellenberg2011}. From this one might conclude that in Al$_2$O$_3$ losses of Eu due to oxidation occur which lead to Eu deficiency in the sample. This was referred to as "intrinsic" p-type doping of EuCd$_2$As$_2$ by \cite{Nelson2024}. Additionally, Al might have entered the material and caused a reduction of the lattice parameters.

Experiments to grow {\bf EuCd$_2$As$_2$ single crystals using Sn flux} were performed by many authors \cite{Gati2021, Wang2022, Santos2023, Shi2024, Roychowdhury2023, Kuthanazhi2023, Nelson2024} and yielded comparable $a$ parameters, Fig.~\ref{fig:5a} (open and closed circles), lying between the "ideal" Ta case \cite{Schellenberg2011} and the Al$_2$O$_3$ synthesis with reduced $a$ parameter \cite{Artmann1996}. In one case \cite{Roychowdhury2023}, the $a$ parameter almost matches the "ideal" one. From this one might conclude that Sn protects Eu from crucible contact and oxidation. Since this is a random process there is some spread in Eu losses and also in $a$.
A Sn-flux-growth study using Na doping performed by Kuthanazhi {\it et al.} \cite{Kuthanazhi2023} proposed that Na occupies the Eu sites and showed that Na doping can slightly enhance $a$ Fig.~\ref{fig:5a} (orange circles).
Experiments to grow {\bf EuCd$_2$As$_2$ single crystals from  NaCl/KCl flux} were mainly performed in SiO$_2$ \cite{Wang2016, Rahn2018, Jo2020, Sanjeewa2020} and in one study in Al$_2$O$_3$ crucibles \cite{Roychowdhury2023}.
The combination of NaCl/KCl flux, SiO$_2$ and Eu might lead to an attack of the SiO$_2$ ampule by the NaCl/KCl mixture in which Na and K act as glass softener and by Eu which tends to oxidize. This would lead to loss of Eu in the process (formation of Eu$_2$O$_3$) and pollution of the melt by Na, K, Si which can act as dopants when included in the crystals.
In \cite{Jo2020}, an Eu deficiency and a  strongly reduced $a$ parameter, Fig.~\ref{fig:5a} (orange square), were found while an enhanced $a$ (red square) and no Eu deficiency could be detected in \cite{Sanjeewa2020}.
The differently enhanced $a$, Fig.~\ref{fig:5a} (red square), in \cite{Sanjeewa2020} and also (dark yellow square) in \cite{Rahn2018} can by explained by different amounts of Eu losses/Si doping partially compensated by Na doping.
Furthermore, in Al$_2$O$_3$, the Eu loss can be compensated by adding an Eu excess into the initial melt \cite{Jo2020} which also leads to a larger $a$ parameter, Fig.~\ref{fig:5a} (orange asterisks).
Interestingly, the $a$ parameter reported by \cite{Roychowdhury2023} who used the combination NaCl/KCl flux and  Al$_2$O$_3$ crucible is quite large, Fig.~\ref{fig:5a} (green square), from which one might speculate that Na doping occurred and at the same time Eu losses were reduced (due to avoidance of salt-SiO$_2$ contact). 
To study the intrinsic properties of semiconducting EuCd$_2$As$_2$, the future task will be to grow single crystals with a minimal amount of unintended dopants and/or vacancies which can be achieved only through usage of highest purity elements (already demonstrated in \cite{Santos2023, Kuthanazhi2023}) in combination with an inert crucible material like Ta. To compensate for Eu losses, a slight Eu excess in the initial melt might be advantageous.

Poly crystalline {\bf EuCd$_2$P$_2$} prepared in Ta crucibles \cite{Schellenberg2011} also here show the largest unit cell volume. 
For the CMR material Sn-flux-grown AFM EuCd$_2$P$_2$, strong sample dependencies in the transport behaviour \cite{Wang2021, Usachov2024, Zhang2023} as well as of the electronic structure as determined by ARPES \cite{Zhang2023, Usachov2024} are reported. Some samples exhibit metallic behaviour at low temperatures while others show an increase of the resistivity at low temperatures. 
Above the \neel temperature, due to a possible polaron formation a peak in the resistivity occurs which shows different peak temperatures for samples from different sources ($T^{\rho}=18$\,K \cite{Wang2021, Sunko2023, Chen2024_Cd}, $T^{\rho}=14$\,K \cite{Usachov2024, Zhang2023}). The preparation of samples with the lowest amount of doping is crucial to study the possibly polaron-related effects in greater detail. 
Poly crystalline and Sn-flux grown {\bf EuZn$_2$P$_2$} using Al$_2$O$_3$ crucibles \cite{Krebber2023, Chen2024_Zn, Rybicki2024} show comparable lattice parameters at room temperature. Nevertheless sample dependencies in transport measurements concerning metallicity are reported for this material as well \cite{Krebber2023, Rybicki2024} and growth of single crystals from ultra pure elements in an inert crucible for further studies of the CMR effect is a task for the future.

Poly crystalline {\bf EuCd$_2$Sb$_2$}/{\bf EuZn$_2$Sb$_2$} prepared in Ta \cite{Schellenberg2011}\cite{Schellenberg2010} and C \cite{Zhang2010}\cite{Zhang2008} has similar lattice parameters that are slightly larger than those from Al$_2$O$_3$ crucibles \cite{Artmann1996}\cite{Kluefers1980, Weber2005} which hints to the presence of Eu vacancies or doping with smaller elements like Al. 
Vapour transport \cite{Soh2018, Su2020} yielded AFM EuCd$_2$Sb$_2$ samples with strongly deviating lattice parameters. This is in contrast to the sister compound EuCd$_2$As$_2$ where similar changes in the lattice parameters in some cases were observed for FM samples.
The other EuZn$_2$Sb$_2$ growth experiments that are reported were performed from a stoichiometric melt \cite{Weber2006} or using a self flux (ZnSb \cite{May2012} or Sb \cite{Singh2024}) to reduce side phases. Growth from self-flux can result in crystals with slightly different composition due to a possible homogeneity range of the material therefore also lattice parameters might slightly change.  

%
\section{Methods to increase purity}
As Eu tends to oxidize, a possible loss can be compensated by using an Eu excess in the initial melt instead of the stoichiometric ratio of the elements.
The choice of the crucible material is crucial to reduce Eu losses. Oxide crucible materials (Al$_2$O$_3$, SiO$_2$) can lead to oxidation of Eu at elevated temperatures accompanied by a pollution of the melt (Al, Si). 
While Al$_2$O$_3$ and SiO$_2$ are inert towards Sn, SiO$_2$ is not inert towards salts or salt mixtures (NaCl, NaCl/KCl). Small atoms like Na and K act as mineralizers and glass softeners and may penetrate into the crucible material leading to a decrease of its structural integrity. The effect gets stronger in case of long exposure times at elevated temperatures and can lead to release of Si and oxygen to the melt.
Tantalum and carbon are inert towards the melt at the temperatures used here and a pollution of the melt is less likely.
In any case, the pollution of the melt can be reduced by using the lowest possible maximum temperature, $T_{\rm max}$, in combination with the shortest possible exposure time, $t_{\rm hold}$, at high temperatures. In this view, using a lower elements-to-flux ratio might be advantageous to decrease the working point.
Typically, in crystal growth experiments slow cooling rates are used to enable the growth of crystals of high purity and less defects. In case of Eu compounds, slow growth rates can be disadvantageous due to the tendency of Eu to evaporate and/or oxidize.
Another source for unwanted oxygen intake might occur in case salt or salt mixtures usage. These need to be dried beforehand the crystal growth experiment as they might contain a certain amount of residual water (oxygen).
Usage of ultra pure starting elements (Eu 4N or higher; $T,Pn$ 5N or 6N, and Sn at least 4N) is crucial to minimize unintended doping.\\

\section{Summary}
The review of growth conditions and structural parameters suggests that the change of the lattice parameters can be understood in terms of positive/negative chemical pressure due to minor amounts of  doping resulting from impurities of the starting materials, the crucible material or the flux.  
So far, EuCd$_2$P$_2$, EuZn$_2$P$_2$, EuCd$_2$As$_2$ and EuZn$_2$As$_2$ that show ferromagnetism were  achieved through salt-flux usage during crystal growth. The comparison shows that this, at least for EuCd$_2$As$_2$, resulted a wide spread of lattice parameters for samples from different sources often connected to a high carrier concentration.
To study the intrinsic properties of Sn/salt flux-grown   Eu$T_2Pn_2$, the future task will be to grow single crystals with a minimal amount of unintended dopants and/or vacancies which can be achieved only through usage of highest purity elements in combination with an inert crucible material like Ta. A slight Eu excess in the initial melt might be advantageous to compensate for Eu losses.

\begin{acknowledgments}
The author thanks S. Krebber for valuable discussions and acknowledges funding by the Deutsche Forschungsgemeinschaft (DFG, German Research Foundation) via the SFB/TRR 288 (422213477, project A03). 

\end{acknowledgments}
%






\bibliography{CrystGrowthEu}
\bibliographystyle{apsrev4-2}

\end{document}